\documentclass[times,twocolumn,final]{elsarticle}

\usepackage{cnf}

\usepackage{framed,multirow}

\usepackage{amssymb}
\usepackage{latexsym}
\usepackage{amsmath}

\usepackage{url}
\usepackage{xcolor}
\definecolor{newcolor}{rgb}{.8,.349,.1}

\usepackage{hyperref}

\usepackage[switch,pagewise]{lineno} 

\journal{Combustion \& Flame}

\usepackage{subcaption}
\usepackage{booktabs}
\usepackage{microtype}
\usepackage[version=4]{mhchem}
\usepackage{siunitx}

\begin{document}

\verso{Baykan \& Liu et al.}

\begin{frontmatter}

\title{A convolutional autoencoder and neural ODE surrogate modeling framework applied to transient counterflow flames}%

\author[1]{Mert Yakup \snm{Baykan}\fnref{fn1}}

\author[2]{Weitao \snm{Liu}\fnref{fn1}}
\fntext[fn1]{These authors contributed equally to this work and should be
considered as co-first authors.}

\author[3]{Mohammad Rafi \snm{Malik}}

\author[2]{Thorsten \snm{Zirwes}}

\author[2]{Andreas \snm{Kronenburg}}

\author[3]{Hong G. \snm{Im}}

\author[1]{Dong-hyuk \snm{Shin}\corref{cor1}}
\cortext[cor1]{Corresponding author: Department of Aerospace Engineering,
KAIST, 291 Daehak-ro, Yuseong-gu, Daejeon 34141, Republic of Korea.}
\emailauthor{dhshin@kaist.ac.kr}{Dong-hyuk Shin}

\address[1]{Department of Aerospace Engineering, KAIST, Daejeon,
Republic of Korea}
\address[2]{Institute for Reactive Flows (IRST), University of Stuttgart,
Pfaffenwaldring 31, 70569 Stuttgart, Germany}
\address[3]{Clean Energy Research Platform (CERP), KAUST, Thuwal,
Saudi Arabia}

\begin{abstract}
A novel convolutional autoencoder and neural ODE (CAE–NODE) framework is proposed for a reduced-order model (ROM) applied to transient 2D counterflow flames, as an extension of AE–NODE methods in homogeneous reactive systems to spatially resolved flows. The multidimensional thermochemical fields (256 × 256 grid, 21 variables) obtained from direct numerical simulations (DNS) are used in training the CAE, where convolutional layers learned the underlying spatial correlations, allowing the CAE to construct an unsupervised 3D latent manifold that is physically meaningful, smooth, and continuous in time. This results in a compression ratio of over 400,000 times. The NODE then subsequently learns the continuous-time dynamics on the latent manifold, enabling the prediction of the full temporal evolution of the flames by integrating forward in time from an initial condition. The results demonstrate that the CAE-NODE can accurately capture the entire transient process, including ignition, flame propagation, and the gradual transition to a non-premixed condition, with excellent agreement with the DNS, while adhering to conservation principles at virtually no computational cost compared to the reference DNS. Predictions remain accurate at strain rates outside the training range. Moreover, despite being unsupervised, the learned latent manifold is highly correlated with the flame-state descriptors such as the progress variable, mixture fraction, and the scalar dissipation rate. This study, for the first time, highlights the potential of CAE-NODE for surrogate modeling of unsteady dynamics of multi-dimensional reacting flows.
\end{abstract}


\end{frontmatter}


\section*{Novelty and significance statement}
The novelty of this work is the demonstration that an unsupervised low-dimensional representation of spatially resolved transient counterflow flames, learned by a data-driven CAE-NODE framework, organizes itself according to physically meaningful flame-state variables. Despite never being prescribed during training, the learned three-dimensional manifold is well described by the domain-averaged mixture fraction, scalar dissipation rate, and conditional progress variable at stoichiometry. The significance of this result is that the coordinates that emerge are closely related to those used in flamelet-generated manifold models and organize the entire temporal evolution, including the ignition and propagation stages. Because the coordinates are recovered rather than assumed, the same framework can potentially be used to identify the reduced coordinates that govern other transient flames in configurations where no established manifold formulation exists and the appropriate variables are not known in advance.

\section{Introduction}
\label{sec1}
As the world moves toward a more sustainable future, the use of alternative fuels such as hydrogen and ammonia has grown significantly~\cite{BLAKEY20112863, CHEHADE2021120845}. In aerospace applications, methane has recently been identified as an attractive fuel for reusable rocket engines, particularly for booster stages~\cite{BurkhardtJSR}, and is now employed by SpaceX in the Raptor engine that powers Starship. Sustainable aviation fuels (SAF) are actively investigated for future aviation applications. Clearly, combustion of hydrocarbon and renewable fuels remains important for energy systems far into the future, and high-fidelity predictive simulations serve as an efficient tool for design and optimization.

The main computational burden in predictive simulations of reacting flows with detailed chemistry is their high dimensionality, especially with detailed chemical mechanisms, and numerical stiffness. As an attempt to develop high-fidelity reduced-order models (ROM), physics-based reductions using mixture fraction or progress variable have limited applicability at conditions outside the validity of the underlying physical approximations \cite{zdybal2025, SHIN20192215}. Data-driven ROMs offer greater flexibility by learning compact representations without presumed intuition. For homogeneous reactive systems, Vijayarangan et al. \cite{vijayarangan2024data} introduced an autoencoder-neural ODE (AE–NODE) framework, where the autoencoder extracts a low-dimensional manifold of thermochemical states, and the neural ODE models their temporal evolution. This method achieved accurate predictions for zero-dimensional (0D) \ce{H2} and \ce{C2H4} autoignition and was later extended to \ce{CH4} \cite{kumar2024b} and \ce{NH3}/\ce{H2} \cite{Baykan25092025} flames. 

For realistic multi-dimensional systems, the AE-NODE approach faces limitations. As a test configuration, counterflow non-premixed flames have been commonly used for a systematic investigation of the flow-chemistry interaction, such as ignition and extinction characteristics \cite{Lutz1997OPPDIFAF,IM01092000}. More realistic two-dimensional effects in such a canonical flame configuration have subsequently been investigated \cite{BURRELL20171387}.
Liu et al. \cite{liu2025, LIU2026100445} examined the performance of AE in two-dimensional laminar counterflow flames, showing their potential but also revealing stability challenges likely due to the limitations of the data set and the lack of spatial awareness in standard AEs. In the transport-reaction systems, mappings between the physical and latent spaces must be performed at every time step. This repeated encoding and decoding introduces additional reconstruction errors, which can hinder the applicability of AE–NODE frameworks beyond 0D simulations.

As a remedy, convolutional autoencoders (CAEs) are naturally suited for spatially structured fields, where convolutional layers capture spatial correlations~\cite{masci2011stacked}. Lee and Carlberg~\cite{lee2020model} showed that CAEs can learn nonlinear manifolds for Galerkin projections, outperforming linear proper orthogonal decomposition (POD)–based approaches in advection-dominated systems, which is a standard ROM technique in combustion applications ~\cite{sutherland2009combustion, isaac2014reduced, malik2023dimensionality}.  It is well known that a shallow linear autoencoder recovers the same principal subspace as POD ~\cite{milano2002neural}, though without descending order of variance unless additional orthogonalization is applied ~\cite{plaut2018principal}.  CAE-based models have also been applied to combustion simulations, demonstrating accurate reconstruction of spatio-temporal flame fields and improved performance over POD-based ROMs~\cite{ZAPATAUSANDIVARAS2024105382}; however, a surrogate model capable of capturing continuous-time dynamics and generalizing across varying operating conditions has not yet been demonstrated.

Building on these ideas, the objective of this study is to develop and evaluate a data-driven reduced-order model capable of accurately predicting the full transient evolution of spatially resolved counterflow flames, from ignition to the transition to a non-premixed flame, while significantly reducing the computational cost compared to high-fidelity simulations. To achieve this, a novel CAE–NODE architecture is proposed as a simulation surrogate for two-dimensional transient counterflow methane/air flames spanning multiple combustion regimes. The model performance is assessed in terms of flame dynamics, prediction accuracy, temporal evolution in the latent space, and generalization to unseen strain rates.

\section{Methodology}

\subsection{Flame configuration\label{subsec:subsection}} \addvspace{10pt}

Two-dimensional (2D) transient laminar counterflow methane–air flames under atmospheric conditions are simulated using the DRM19 chemical mechanism \cite{DRM19}. The simulations are performed by a fully compressible Navier--Stokes solver developed within the framework of OpenFOAM~\cite{jasak2007openfoam}.

The initial species composition is set from the steady-state condition of the non-reacting counterflow simulation. Ignition is initiated by a hot spot of size 0.9 mm × 1.0 mm with a temperature of 1950 K. After the flame kernel is formed, the flame propagates away from the centerline, gradually transitioning into a steady non-premixed flame. The surrogate model (CAE-NODE) introduced in this study is designed to capture the entire 2D temporal evolution, from ignition to kernel formation to transition to a diffusion flame. 

Exemplary snapshots of simulations are shown in the left columns of Figs. \ref{snapshots_SR40}, and \ref{fig:snapshots_unseen_combined}. The domain consists of 256 × 256 cells with a uniform cell size of \SI{78.125}{\mu\meter} spanning over 20×20 mm$^2$. The fuel of pure methane flows in from the top, and the oxidizer flows in from the bottom, comprising air (23\% \ce{O2} and 77\% \ce{N2} by mass fraction), both at an inlet temperature of T$_0=\SI{650}{\kelvin}$. The simulated global strain rates (SR) are SR $\in$ [6, 8, 10, 15, 25, 40, 80, 120] $\mathrm{s}^{-1}$ for training and SR $\in$ [4, 12, 20, 22, 30, 60, 100, 140, 160] $\mathrm{s}^{-1}$ for validation.

\subsection{Surrogate modelling: CAE-NODE \label{subsec:subsection}} 
\begin{figure*}[!ht]
\centering
\includegraphics[width=400pt]{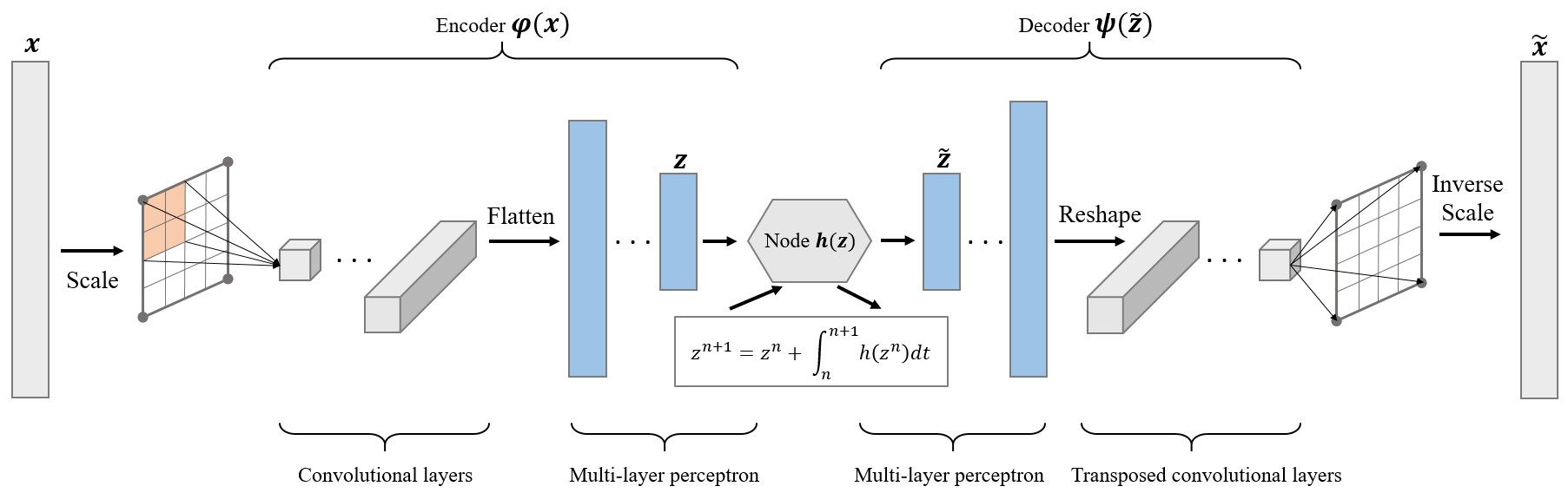}
\vspace{0 pt}
\caption{\footnotesize Combined CAE and NODE architecture.}
\label{architecture}
\end{figure*}

\addvspace{10pt}
Figure~\ref{architecture} shows the overall architecture of the surrogate model. During inference, the ROM employs a CAE to compress the initial thermochemical state tensor, $x_0 \in \mathbb{R}^{256\times256\times21}$, where $256\times256$ is the spatial resolution, and $21$ denotes the thermochemical state variables consisting of the 20 species mass fractions from the DRM19 mechanism (excluding argon \ce{Ar}) and temperature ($Y_{\ce{H2}}, \ldots, Y_{\ce{N2}}, \text{T}$), into a 3-dimensional latent vector, $\varphi(x_0) = z_0 \in \mathbb{R}^{3}$, through nonlinear convolutional encoding. Then, the NODE is used to integrate $z_0$ in time to obtain the latent trajectory, $\text{ODESolve}(z_0, h_\theta, t_{1, \ldots, 500}) = \tilde{z} \in\mathbb{R}^{500\times3}$, which is then decoded to reconstruct the physical fields, $\psi(\tilde{z}) = \tilde{x} \in \mathbb{R}^{500\times256\times256\times21}$, where $500$ is the number of temporal snapshots. Note that, since the proposed CAE-NODE framework is designed as a surrogate for the entire simulation, the encoder is used only once during inference to map the initial fields from physical space to latent space, while the decoder is required only to reconstruct the physical states for post-processing. This strategy prevents reconstruction errors from propagating through the temporal predictions.

The convolutional architecture follows the design proposed by Lee and Carlberg~\cite{lee2020model}, in which spatial compression is performed exclusively by learned strided convolutions. The encoder consists of four convolutional layers using $5\times5$ kernels with half-padding and a stride of 2, so that each layer halves the spatial resolution and the input $256\times256$ cells of the computational domain are systematically reduced to a $16\times16$ grid at the bottleneck, corresponding to a fixed $1/16$ convolutional compression ratio per spatial dimension. The channel schedule expands by a factor of 16 relative to the input: starting from the 21 (state vector) input channels, the filter count doubles across successive layers, progressing from 42 to 84, 168, and finally 336 filters at the deepest convolutional level. The resulting $16\times16\times336$ feature map is flattened into an $86{,}016$-dimensional vector and mapped directly to the latent space through a single dense layer. The decoder mirrors this construction exactly, projecting the latent variables back to the flattened bottleneck dimension through a single dense linear layer, after which four transposed convolutional layers with stride 2 restore the spatial resolution, each doubling the grid size until the original $256\times256\times21$ field is recovered. The NODE consists of five fully connected layers with 200 neurons each, using ELU activations except for the output layer. Layer details are summarized in Table~\ref{tab:CAE+NODE Layers}. The number of fully connected layers for the NODE and the number of neurons per layer were chosen to be similar to those used in previous studies \cite{vijayarangan2024data, Baykan25092025}, ensuring sufficient model capacity while avoiding excessive complexity that could lead to overfitting.

\begin{table}[h]
\footnotesize
\setlength{\tabcolsep}{3pt} 
\renewcommand{\arraystretch}{0.9} 
\caption{Convolutional Autoencoder architecture.}
\centering
\resizebox{1\columnwidth}{!}{%
\begin{tabular}{cc}
\begin{tabular}{|c|c|}
\hline
\multicolumn{2}{|c|}{\textbf{Encoder network}} \\
\hline
\multicolumn{2}{|c|}{\textit{Convolutional layers}} \\
\hline
\textbf{Layer} & \textbf{Filters} \\
\hline
1 & 42 \\
2 & 84 \\
3 & 168 \\
4 & 336 \\
\hline
\multicolumn{2}{|c|}{\textit{Fully-connected layers}} \\
\hline
\textbf{Layer} & \textbf{Input $\to$ Output} \\
\hline
1 & 86016 $\to$ 3 \\
\hline
\end{tabular}
&
\begin{tabular}{|c|c|}
\hline
\multicolumn{2}{|c|}{\textbf{Decoder network}} \\
\hline
\multicolumn{2}{|c|}{\textit{Fully-connected layers}} \\
\hline
\textbf{Layer} & \textbf{Input $\to$ Output} \\
\hline
1 & 3 $\to$ 86016 \\
\hline
\multicolumn{2}{|c|}{\textit{Transposed convolutional layers}} \\
\hline
\textbf{Layer} & \textbf{Filters} \\
\hline
1 & 168 \\
2 & 84 \\
3 & 42 \\
4 & 21 \\
\hline
\end{tabular}
\\
\end{tabular}
} 
\label{tab:CAE+NODE Layers}
\end{table}

\subsection{Training procedure}

The CAE and NODE are trained jointly for 2,000 epochs (approximately 12 hours) using a hybrid optimization strategy. The training dataset consists of eight transient flame simulations, each containing 500 snapshots, for a total of 4,000 snapshots. During each epoch, the reconstruction losses ($L_1$ and $G_1$) update the CAE parameters, the projection loss ($L_3$) updates the NODE parameters, and the prediction loss ($L_2$) updates the combined CAE-NODE parameters. The latent dimension chosen for this study is 3, as we observed that the accuracy of the model does not change significantly with higher latent dimensions. The NODE models continuous latent dynamics conditioned on the logarithm of the strain rate, resulting in an augmented latent state of dimension $3+1$. The optimization procedure uses PyTorch and the Adam optimizer, and is detailed below \cite{paszke2019pytorch, kingma2014adam}.

\subsubsection{Snapshot-level reconstruction (L1 + G1)}
The CAE is trained on random snapshots drawn uniformly from 70\% of the 4,000 training snapshots (2,800). The run processes these in mini-batches of 4, resulting in the CAE parameters being updated 700 times per epoch. This stream calculates the reconstruction loss ($L_1$), and an additional first-order spatial gradient loss ($G_1$) to better capture the profiles of minor species:
\begin{equation}
L_1 = ||\psi(\varphi(x)) - x||_{2}^{2}, \label{eq:L1}
\end{equation}
where $\varphi$ is the encoder and $\psi$ is the decoder.

\begin{equation}
G_1 = ||\nabla_x (\psi(\varphi(x)) - x)||_2^2 + ||\nabla_y (\psi(\varphi(x)) - x)||_2^2 \label{eq:G1}
\end{equation}

\subsubsection{Trajectory-level NODE integration (L2 + L3)}
Simultaneously, the NODE tracks full temporal trajectories. The prediction loss ($L_2$), which evaluates the physical-space error of the NODE forecast, is computed 8 times per epoch (once for each strain rate simulation). Additionally, an inner loop is executed where the projection loss ($L_3$), which evaluates NODE's time-evolution accuracy strictly within the manifold, is computed and the NODE parameters are updated 50 times.

\begin{equation}
\begin{gathered}
L_2 = \left|\left| \psi(\tilde{z}^{(m)}_{1, \ldots, 500}) - x^{(m)}_{1, \ldots, 500} \right|\right|_2^2, \label{eq:L2} \\ \\
\end{gathered}
\end{equation}
\begin{equation}
\begin{gathered}
L_3 = \left|\left| \tilde{z}^{(m)}_{1, \ldots, 500} - \varphi(x^{(m)}_{1, \ldots, 500}) \right|\right|_2^2,  \label{eq:L3} \\ \\
\tilde{z}^{(m)}_{1, \ldots, 500} = \text{ODESolve}(z_0 = \varphi(x^{(m)}_0), h_\theta, t_{1, \ldots, 500}),
\end{gathered}
\end{equation}
where $1 \leq m \leq 8$ refers to the $m^{th}$ strain rate simulation, and where $h_\theta$ refers to NODE parameters.$x_0$ and $z_0$ refer to the state vector and the latent vector at the initial condition. Subscript ${1, \ldots, 500}$ denotes time steps from 1 to 500 with an increment of 1.

\subsubsection{Combined objective}
The total objective function optimized during the hybrid joint training merges the snapshot and trajectory-level loss functions, with the spatial gradient loss weighted at $10^2$ to match the scales of the other loss functions:
\begin{equation}
L = L_1 + 10^2 \cdot G_1 + L_2 + L_3
\end{equation}

\section{Results}

In this section, we analyze and describe the performance of the CAE-NODE model with respect to the reference DNS and provide a comparison with the widely used principal component (PC)-transport model. The model's performance is evaluated in terms of its accuracy, adherence to physical principles, and computational gains. The analysis involves five sections. (i) Validation within the training data, (ii) validation for unseen strain rates that are within (interpolation) and out of (extrapolation) the training data distribution, (iii) analysis of the latent manifold, (iv) analysis of stiffness reduction with a computational speedup comparison, and (v)  comparison with PC-transport ROM.

Before presenting the results, the error metrics and the reduced
thermochemical coordinates used throughout this section are defined. Both error
metrics are evaluated separately for each thermochemical variable $k$, so that
the accuracy of the major and minor species can be assessed independently. The
root mean squared error (RMSE) quantifies the absolute deviation between the
DNS field $x$ and the CAE-NODE prediction $\tilde{x}$ in the units of the
variable considered,
\begin{equation}\label{eq:RMSE}
\mathrm{RMSE}_k =
\sqrt{\frac{1}{N_t N_x N_y}
\sum_{n=1}^{N_t}\sum_{i=1}^{N_x}\sum_{j=1}^{N_y}
\left(x_{nijk} - \tilde{x}_{nijk}\right)^2},
\end{equation}
where $N_t = 500$ is the number of temporal snapshots and
$N_x = N_y = 256$ the number of cells in each spatial direction. Since the
thermochemical variables span several orders of magnitude, the RMSE alone does
not allow a comparison of accuracy across species. The relative Frobenius error
(RFE) is therefore introduced as a normalized counterpart,
\begin{equation}\label{eq:Frobenius Error}
\begin{split}
\mathrm{RFE}_k = \frac{\|x_k - \tilde{x}_k\|_F}{\|x_k\|_F}
&= \frac{\sqrt{\sum_{n,i,j}(x_{nijk} - \tilde{x}_{nijk})^2}}
        {\sqrt{\sum_{n,i,j} x_{nijk}^2}}, \\
\tilde{x}_n &= \psi(\tilde{z}_{n}).
\end{split}
\end{equation}
Both measures are reported with two subscripts. The subscript $\mathrm{space}$
denotes evaluation over the full two-dimensional field, i.e.\ the summations in
Eqs.~\eqref{eq:RMSE} and~\eqref{eq:Frobenius Error} are taken over all cells of
the computational domain, so that an error in the physical position of the
flame contributes directly. The subscript $\mathrm{manifold}$ denotes the same
measures evaluated in the thermochemical composition space, where the DNS and
predicted states are compared at equal mixture fraction rather than at equal
physical location. The comparison of the two therefore separates a
displacement of the flame in physical space from an error in the underlying
thermochemical state.
 
In addition to the raw state variables, to evaluate the CAE-NODE framework, three reduced coordinates are used to
characterize the flames and to interpret the learned latent manifold in
Section~\ref{sec:latent}. The progress variable $C$ is defined as a weighted
sum of the mass fractions of the major and intermediate products \cite{vanoijen2016},
\begin{equation}\label{eq:progress_variable}
C = 1.2\,Y_{\ce{CO2}} + 0.9\,Y_{\ce{CO}} + 2.7\,Y_{\ce{HO2}}
  + 1.5\,Y_{\ce{CH2O}} + 1.2\,Y_{\ce{H2O}},
\end{equation}
where the weights are chosen so that $C$ increases monotonically through the
reaction zone for the conditions considered here. The mixture fraction $Z$ is
computed following Bilger~\cite{bilger1989structure}, based on the elemental mass
fractions $Z_\mathrm{C}$, $Z_\mathrm{H}$ and $Z_\mathrm{O}$ and the
corresponding atomic masses $W_\mathrm{C}$, $W_\mathrm{H}$ and $W_\mathrm{O}$,
\begin{equation}\label{eq:bilger}
Z = \frac{\beta - \beta_\mathrm{ox}}{\beta_\mathrm{f} - \beta_\mathrm{ox}},
\qquad
\beta = \frac{2 Z_\mathrm{C}}{W_\mathrm{C}}
      + \frac{Z_\mathrm{H}}{2 W_\mathrm{H}}
      - \frac{Z_\mathrm{O}}{W_\mathrm{O}},
\end{equation}
where the subscripts $\mathrm{f}$ and $\mathrm{ox}$ refer to the fuel and
oxidizer streams, respectively, such that $Z = 1$ in the pure methane stream
and $Z = 0$ in the oxidizer stream. Finally, the scalar dissipation rate
$\chi$, which measures the local rate of molecular mixing and characterizes the
departure from chemical equilibrium, is evaluated from the mixture fraction
gradient as
\begin{equation}\label{eq:chi}
\chi = 2 D \left| \nabla Z \right|^2
     = 2 D \left[
       \left(\frac{\partial Z}{\partial x}\right)^2 +
       \left(\frac{\partial Z}{\partial y}\right)^2
       \right],
\end{equation}
where $D$ is the thermal diffusivity of the mixture under a unity-Lewis number assumption.

\subsection{Validation of the CAE-NODE architecture}

\begin{figure*}[ht!]
\centering
\begin{subfigure}[h]{1.0\textwidth}
    \centering
    \includegraphics[width=\textwidth]{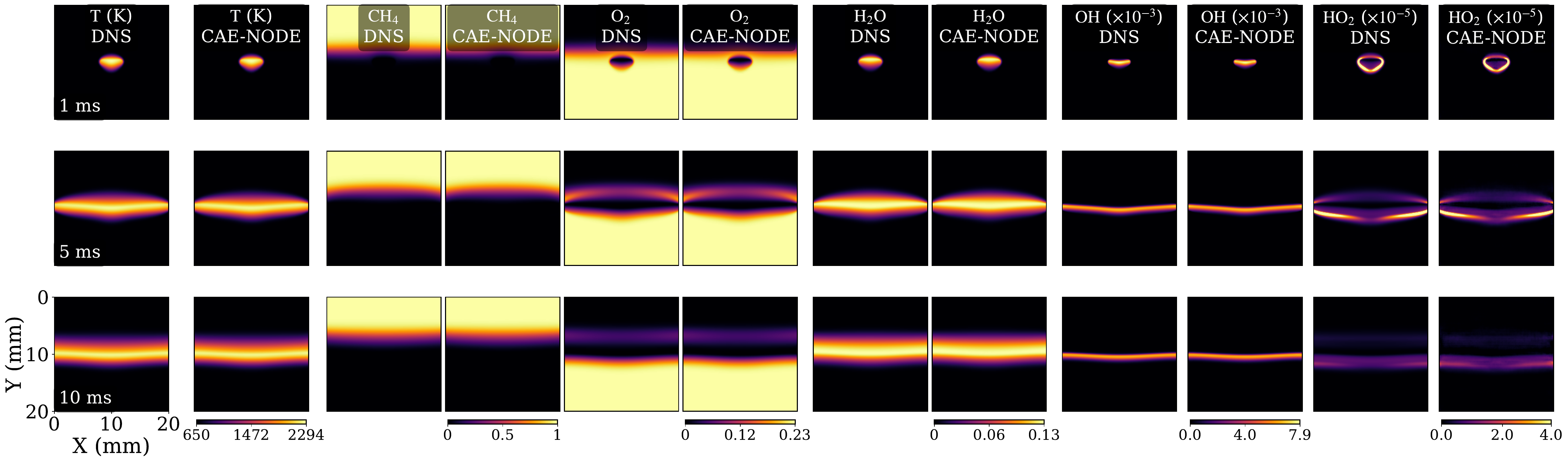}
    \caption{\footnotesize Temperature, and Y$_k$ snapshots.}
    \label{snapshots_SR40}
\end{subfigure}
\begin{subfigure}[h]{0.9\textwidth}
    \centering
    \includegraphics[width=\textwidth]{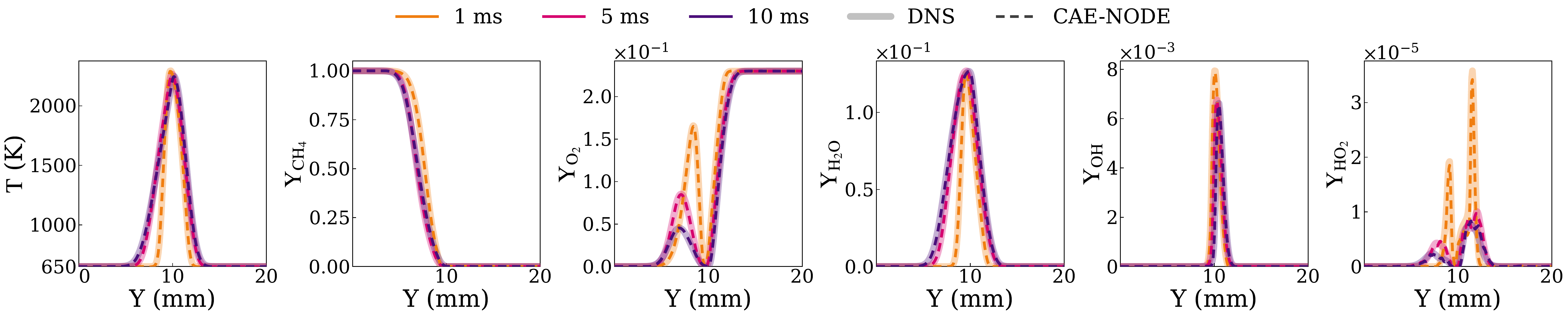}
    \caption{\footnotesize Temperature, and Y$_k$ along the centerline.}
    \label{centerline_SR40}
\end{subfigure}
\begin{subfigure}[h]{0.9\textwidth}
    \centering
    \includegraphics[width=\textwidth]{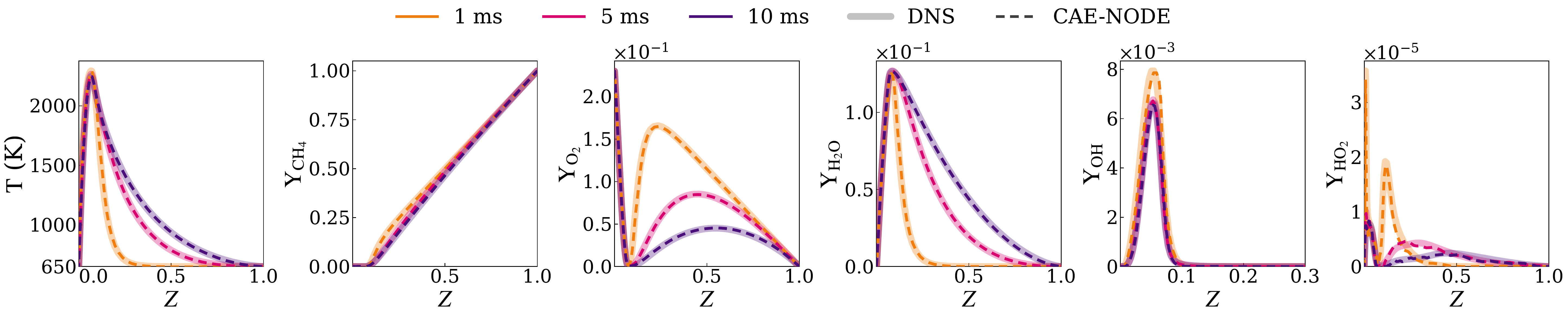}
    \caption{\footnotesize Temperature, and Y$_k$ vs mixture fraction ($Z$) along the centerline.}
    \label{centerline_Zspace_SR40}
\end{subfigure}

\caption{\footnotesize Comparison of DNS vs. CAE-NODE for SR = $40~s^{-1}$.}
\label{fig:center_combined_SR40}
\end{figure*}

Figure~\ref{snapshots_SR40} compares the DNS and CAE-NODE snapshots of temperature and selected species mass fractions at 1, 5, and 10\,ms for
SR~$=40\,\mathrm{s}^{-1}$, a strain rate included in the training set. The three instants correspond to the three stages of the transient flame development: the ignition
kernel formed by the hot spot at 1\,ms, the lateral propagation of the flame away from the centerline at 5\,ms, and the fully established, quasi
one-dimensional diffusion flame at 10\,ms. Starting with the same initial condition as the DNS as input, the surrogate model predicts the next solution of the full state variable vector in all 256x256 cells at 500 time instances throughout all flame development stages, successfully reproducing the flame's evolution, capturing all
three stages, including the slight curvature of the propagating front at 5\,ms and its subsequent flattening into a horizontal reaction layer. The model produces accurate results for minor species as well.
The thin \ce{OH} layer that marks the reaction zone and the annular \ce{HO2} structure surrounding the ignition kernel at
1\,ms are both predicted with the correct shape, magnitude and spatial location.
 
A more quantitative assessment is provided by the centerline profiles in Fig.~\ref{centerline_SR40}. Temperature, Y$_{\ce{CH4}}$, Y$_{\ce{O2}}$, and Y$_{\ce{H2O}}$ are
essentially indistinguishable from the DNS at all three instants, with the model correctly capturing the narrow, high-temperature kernel at 1\,ms and the
broadening of the profiles as the flame relaxes toward the diffusion-controlled state. The peak location and magnitude of $Y_{\ce{OH}}$ are reproduced
throughout, and even the transient double-peak structure of $Y_{\ce{HO2}}$ at 1\,ms is captured. This behavior is reflected in Table~\ref{tab:err_sr40}: the relative Frobenius error
($\mathrm{RFE}_\mathrm{space}$) remains below 0.5\,\% for temperature and the major species, rises to 1.37\,\% for \ce{OH}, and reaches 11.64\,\% for
\ce{HO2}. The increase in error for the minor species stems primarily from their fields being close to zero across most of the domain, with only a few sharp localized peaks. Because the reconstruction loss (Eq.~\eqref{eq:L1}) is based on the mean squared error (MSE), it inherently minimizes the average error over the field and therefore does not necessarily reproduce these localized peaks accurately. The gradient loss $G_1$ was introduced to limit this effect for the minor species, by forcing the model to give more importance to sharp gradients found in the spatial profiles of the minor species, compared to the smoothly varying profiles of the major species.
 
Because the intended use of the surrogate is the prediction of thermochemical states, it is essential to examine the same data in thermochemical composition space.
Figure~\ref{centerline_Zspace_SR40} shows the centerline quantities plotted against the mixture fraction $Z$. The CAE-NODE prediction preserves the state-space
structure of the flame, the linear $Y_{\ce{CH4}}$-$Z$ relation, the progressive decrease of \ce{O2} on the rich side as the flame develops, and
the peak of $Y_{\ce{OH}}$ near the stoichiometric value. The evolution from the ignition-dominated state at 1\,ms to the established flame
structure at 10\,ms is followed correctly. The corresponding manifold errors in Table~\ref{tab:err_sr40} are of the same order as the
physical-space errors (0.74\,\% for temperature, 9.30\,\% for \ce{HO2}).
 
\begin{figure}[ht!]
    \centering
    \includegraphics[width=120pt]{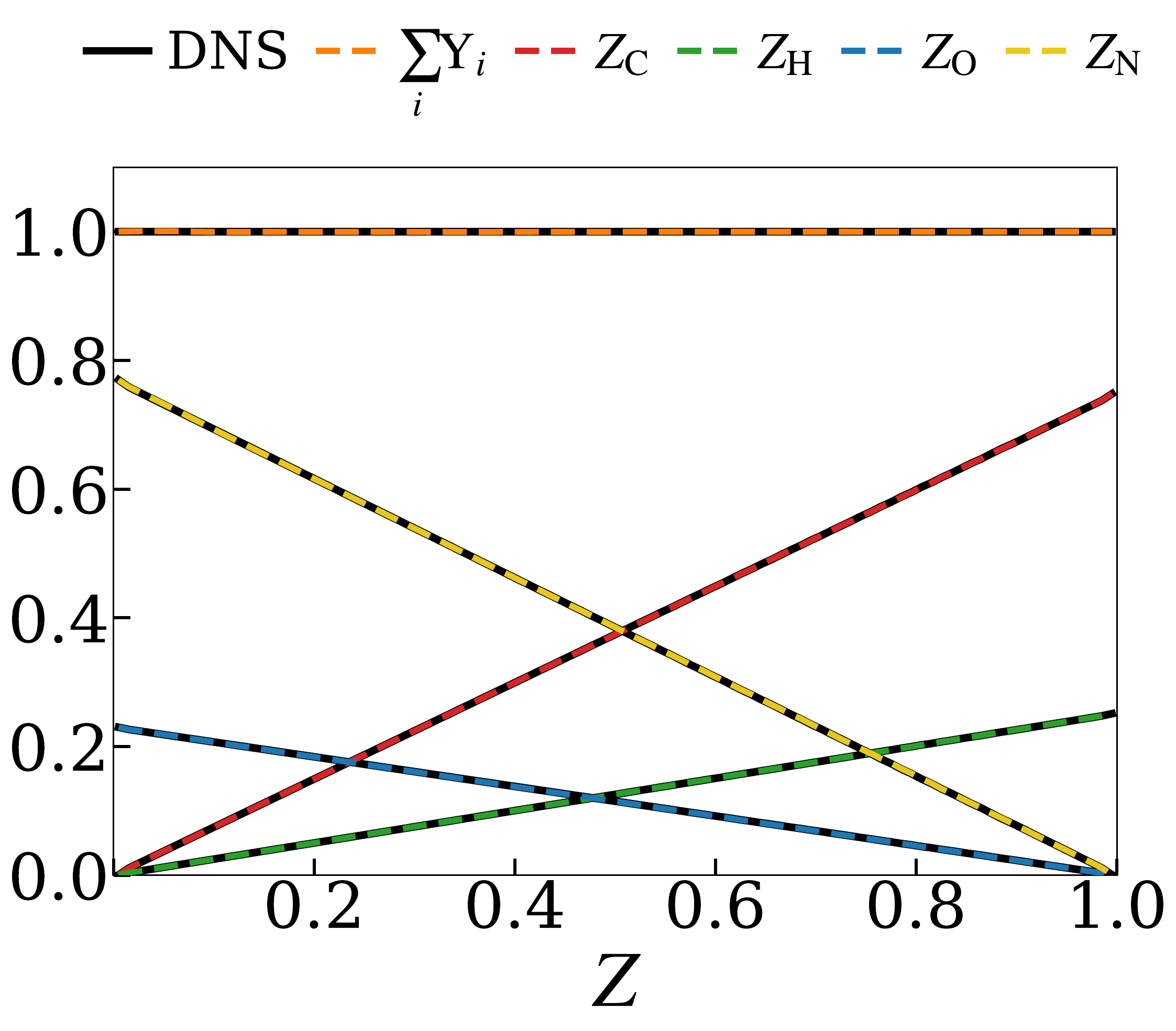}
    \caption{\footnotesize Sum of mass fractions and elemental mass fractions versus mixture-fraction for SR = $40~s^{-1}$.}
    \label{conservatrion_SR40}
\end{figure}

Finally, Fig.~\ref{conservatrion_SR40} examines whether the decoded fields remain physically consistent. The sum of the mass fractions returned by the
decoder is near unity across the whole range of mixture fraction, and the elemental mass fractions $Z_\mathrm{C}$, $Z_\mathrm{H}$, $Z_\mathrm{O}$, and
$Z_\mathrm{N}$ vary linearly between the oxidizer and fuel boundary values, in excellent agreement with the DNS. No constraint enforcing mass or element conservation is
imposed at any point during training, and these properties are learned from the data alone.

\begin{table}[htbp!]
\centering
\caption{RMSE and RFE of CAE-NODE predictions for SR = $40$~s$^{-1}$. Chemical species name represents mass fraction for that species.}
\setlength{\tabcolsep}{4pt}
\footnotesize
\begin{tabular}{lcccc}
\toprule
 & \multicolumn{2}{c}{RMSE} & \multicolumn{2}{c}{RFE [\%]} \\
\cmidrule(lr){2-3} \cmidrule(lr){4-5}
T/Y$_k$ & space & manifold & space & manifold \\
\midrule
T [K]   & 2.72                  & 6.68                  & 0.30 & 0.74 \\
CH$_4$  & $1.34\times10^{-3}$   & $6.49\times10^{-4}$   & 0.23 & 0.11 \\
O$_2$   & $4.23\times10^{-4}$   & $1.10\times10^{-3}$   & 0.26 & 0.69 \\
H$_2$O  & $1.45\times10^{-4}$   & $6.65\times10^{-4}$   & 0.41 & 1.95 \\
OH      & $1.42\times10^{-5}$   & $1.75\times10^{-5}$   & 1.37 & 1.80 \\
HO$_2$  & $7.97\times10^{-7}$   & $3.63\times10^{-7}$   & 11.64 & 9.30 \\
\bottomrule
\end{tabular}
\label{tab:err_sr40}
\end{table}

\subsection{Validation for Unseen Strain Rates}

\begin{figure*}[ht!]
\centering

\begin{subfigure}[h]{1\textwidth}
    \centering
    \includegraphics[width=\textwidth]{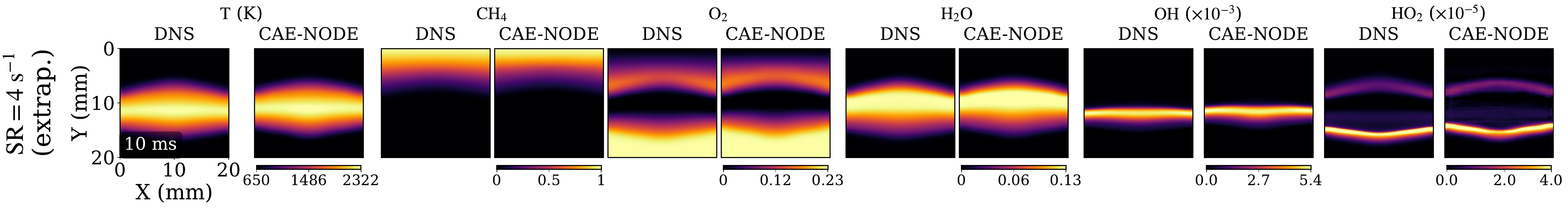}
    \label{snapshots_SR4}
    \vspace{-10pt}
\end{subfigure}
\vspace{-10pt}
\begin{subfigure}[h]{1\textwidth}
    \centering
    \includegraphics[width=\textwidth]{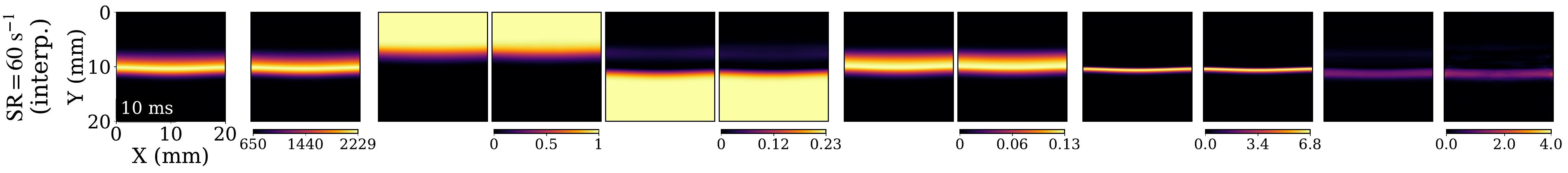}
    \label{snapshots_SR60}
\end{subfigure}
\vspace{-10pt}
\begin{subfigure}[h]{1\textwidth}
    \centering
    \includegraphics[width=\textwidth]{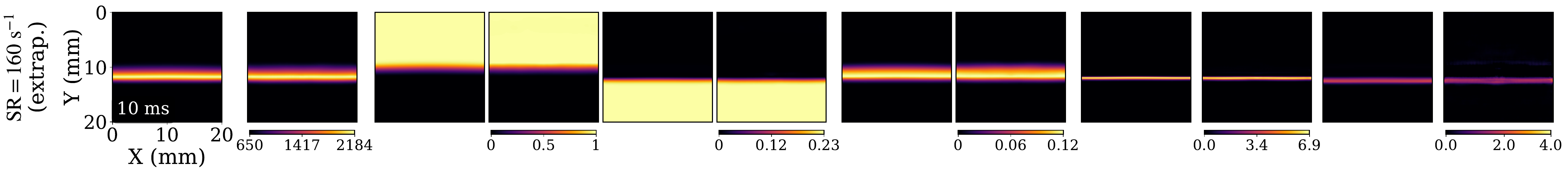}
    \label{snapshots_SR160}
\end{subfigure}

\caption{\footnotesize Temperature, and Y$_k$ snapshots for unseen strain rates, DNS vs CAE-NODE.}
\label{fig:snapshots_unseen_combined}
\end{figure*}

The generalization of the model is assessed for strain rates not seen during training. SR~$=60\,\mathrm{s}^{-1}$ lies within the range of strain rates used for training and is therefore referred to as interpolation, whereas SR~$=4$ and $160\,\mathrm{s}^{-1}$ fall below and above that range and are referred to as extrapolation. Figure~\ref{fig:snapshots_unseen_combined} shows the corresponding fields at 10\,ms. As in the validation for trained conditions, predictions start from the same initial condition as the DNS; the surrogate model predicts the next solution of the full state variable vector in all $256\times256$ cells at 500 time instances spanning all flame development stages. The DNS exhibits the expected response to the mixing intensity (i.e. strain rate for this setup): the reaction layer becomes progressively thinner and flatter as SR
increases, while the peak temperature drops from 2320\,K at SR~$=4\,\mathrm{s}^{-1}$ to 2220\,K at $60\,\mathrm{s}^{-1}$ and 2180\,K at
$160\,\mathrm{s}^{-1}$. The CAE-NODE reproduces this thinning and the associated reduction in peak temperature, together with the changes in the
two-dimensional curvature of the layer at the lowest strain rate, where the flame is thickest and occupies a substantial fraction of the domain.
 
\begin{figure*}[ht!]
\centering
\begin{subfigure}[h]{0.9\textwidth}
    \centering
    \includegraphics[width=\textwidth]{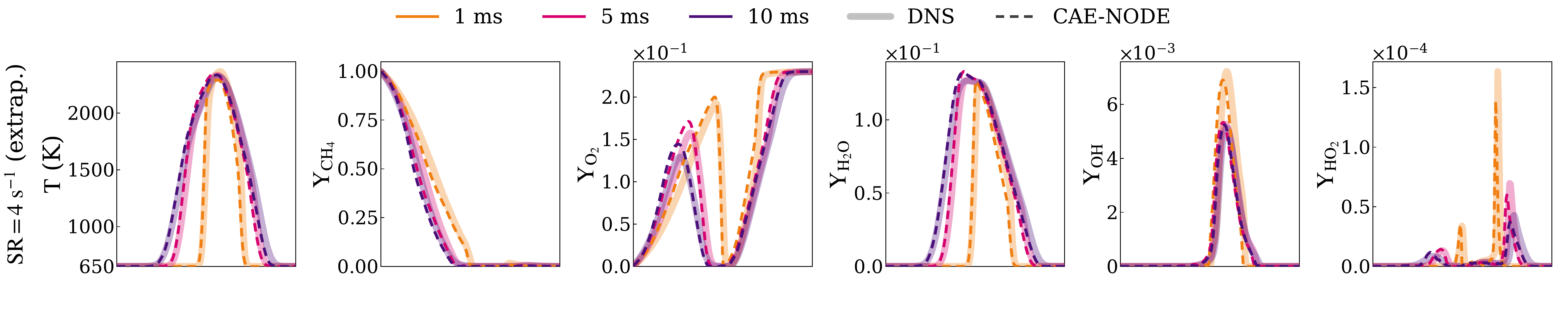}
    \label{centerline_SR4}
    \vspace{-20pt}
\end{subfigure}
\begin{subfigure}[h]{0.9\textwidth}
    \centering
    \includegraphics[width=\textwidth]{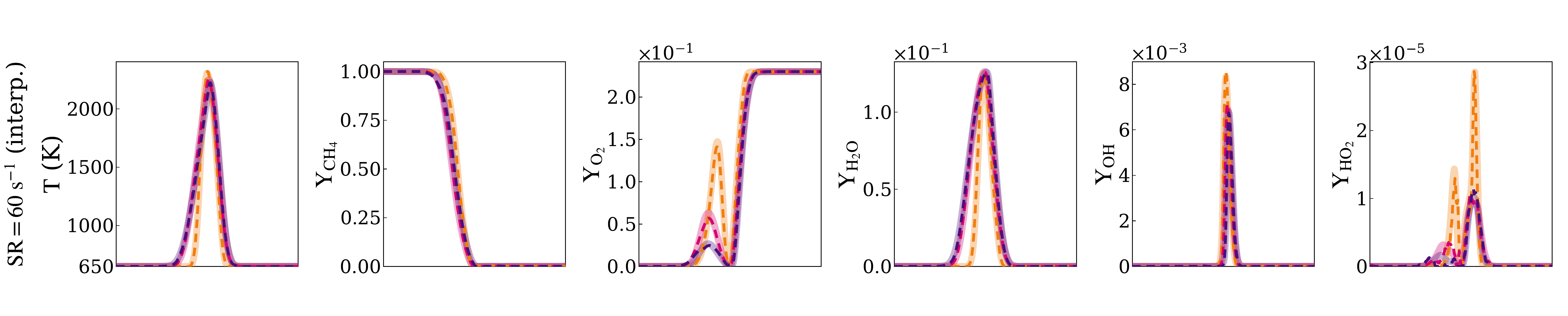}
    \label{centerline_SR60}
    \vspace{-20pt}
\end{subfigure}
\begin{subfigure}[h]{0.9\textwidth}
    \centering
    \includegraphics[width=\textwidth]{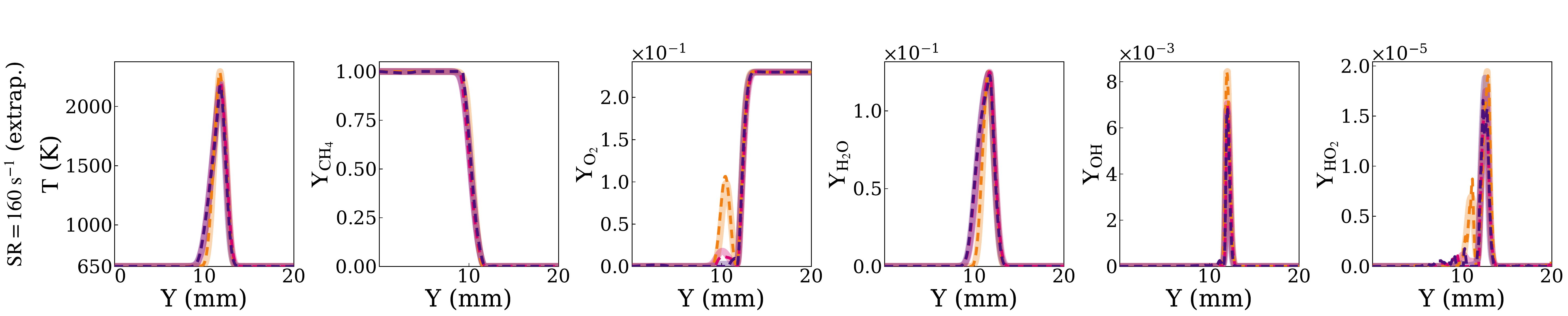}
    \label{centerline_SR160}
    \vspace{-10pt}
\end{subfigure}

\caption{\footnotesize Temperature and Y$_k$ along the centerline for unseen strain rates, DNS vs CAE-NODE.}
\label{fig:center_combined_unseen}
\end{figure*}

The centerline profiles in Fig.~\ref{fig:center_combined_unseen} confirm this behavior.
For the interpolated case, SR~$=60\,\mathrm{s}^{-1}$, the prediction is nearly indistinguishable from the DNS for all variables and all three
instants, and the upper extrapolation at SR~$=160\,\mathrm{s}^{-1}$ is of comparable quality; as shown in Table~\ref{tab:relative_errors_unseen} the errors are 1.71\,\% and 2.74\,\% for temperature, respectively. The lower extrapolation at SR~$=4\,\mathrm{s}^{-1}$ is the most challenging case for the CAE-NODE model as the flame dynamics are more sensitive to the strain rate than at higher strain rates. Despite this, the profile shapes and their magnitudes are predicted with great accuracy, but the location is slightly displaced relative to the DNS.
 
\begin{figure*}[ht!]
\centering
\begin{subfigure}[h]{0.9\textwidth}
    \centering
    \includegraphics[width=\textwidth]{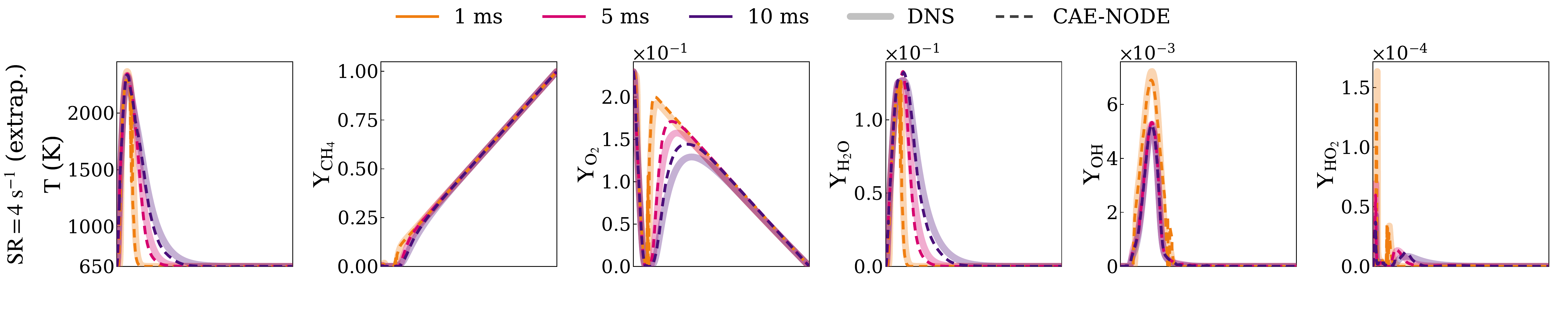}
    \label{centerline_Zspace_SR4}
    \vspace{-20pt}
\end{subfigure}
\begin{subfigure}[h]{0.9\textwidth}
    \centering
    \includegraphics[width=\textwidth]{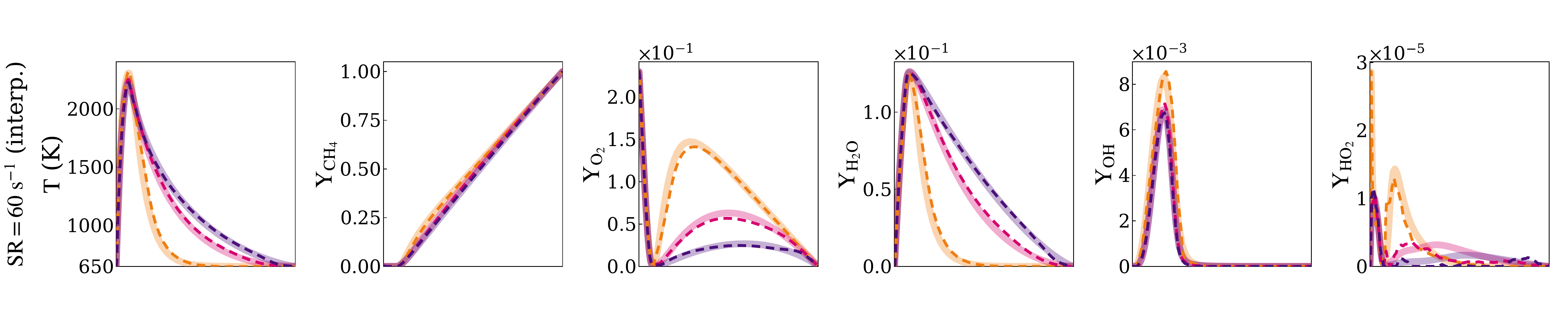}
    \label{centerline_Zspace_SR60}
    \vspace{-20pt}
\end{subfigure}
\begin{subfigure}[h]{0.9\textwidth}
    \centering
    \includegraphics[width=\textwidth]{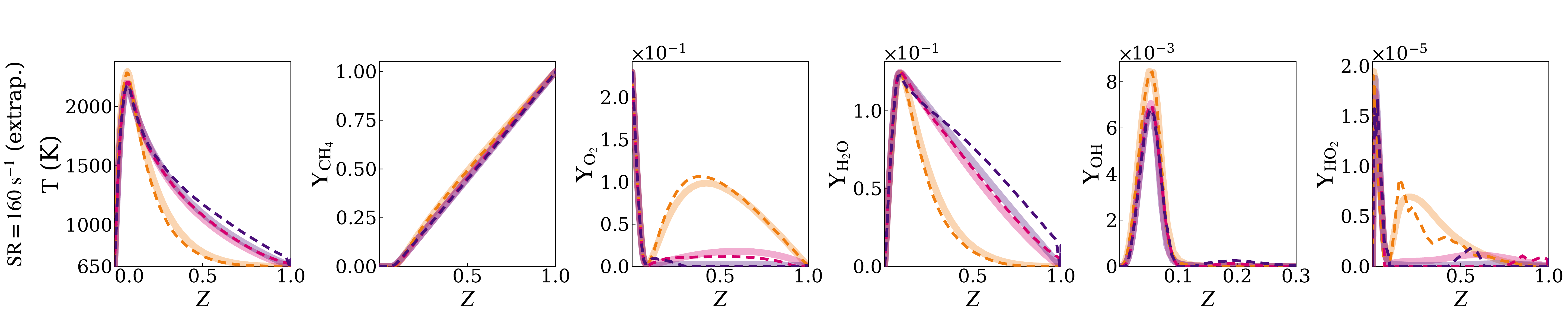}
    \label{centerline_Zspace_SR160}
    \vspace{-10pt}
\end{subfigure}

\caption{\footnotesize Temperature and Y$_k$  vs mixture fraction ($Z$) along the centerline for unseen strain rates, DNS vs CAE-NODE.}
\label{fig:center_Zspace_combined_unseen}
\end{figure*}

\begin{figure*}[ht!]
\centering
\includegraphics[width=0.8\textwidth]{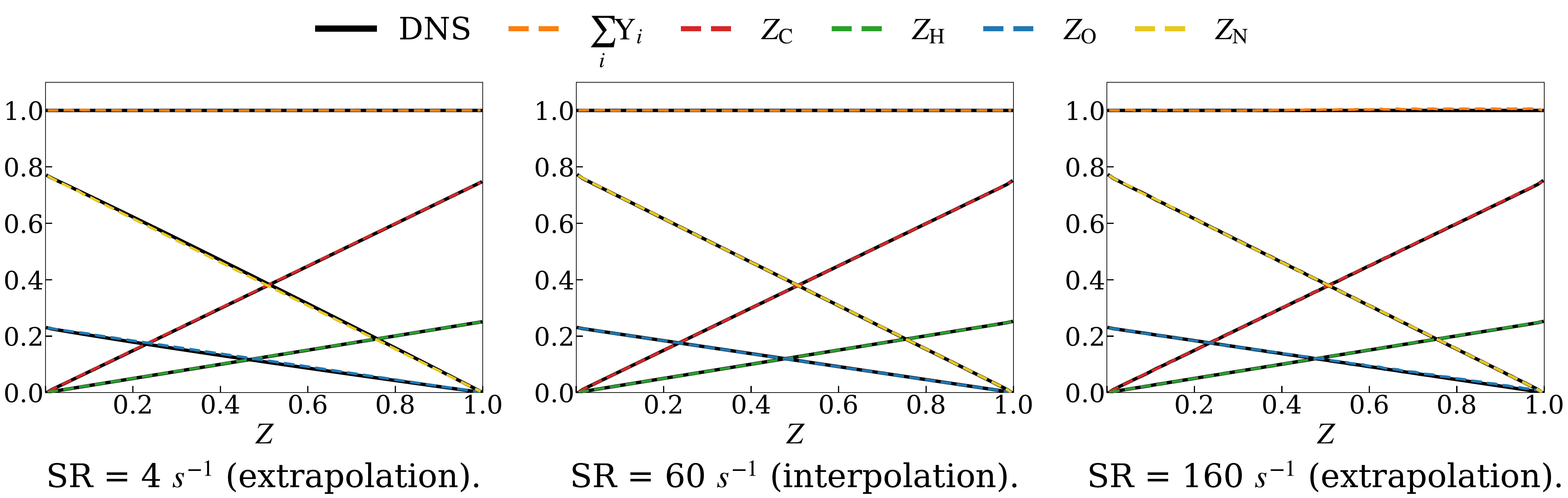}
\caption{\footnotesize Sum of mass fractions and elemental mass fractions versus
mixture-fraction for unseen strain rates.}
\label{fig:conservatrion_combined_unseen_sr}
\end{figure*}

The distinction between a displacement error and a thermochemical error becomes clear when the same data are viewed in the thermochemical composition space. In
Fig.~\ref{fig:center_Zspace_combined_unseen}, the predicted profiles collapse onto the DNS curves for all three unseen strain rates, including SR~$=4\,\mathrm{s}^{-1}$.
This is quantified in Table~\ref{tab:relative_errors_unseen}: for SR~$=4\,\mathrm{s}^{-1}$ the physical-space error is 9.97\,\% for temperature and 103.16\,\% for
\ce{HO2}, whereas the corresponding manifold errors are only 2.13\,\% and 25.63\,\%. The model therefore predicts a flame whose thermochemical structure
is correct but whose position in physical space is shifted; this is consistent with the latent-space behavior discussed in section \ref{sec:latent}. Figure~\ref{fig:conservatrion_combined_unseen_sr} further shows that mass and elemental conservation are satisfied for all unseen strain rates
to the same degree as for the training conditions, i.e.\ the conservation properties of the decoder are not a memorized feature of the training set.
This also indicates that the learned latent manifold accurately captures the underlying physics of the system.
 
\begin{figure*}
    \centering
    \includegraphics[width=0.8\linewidth]{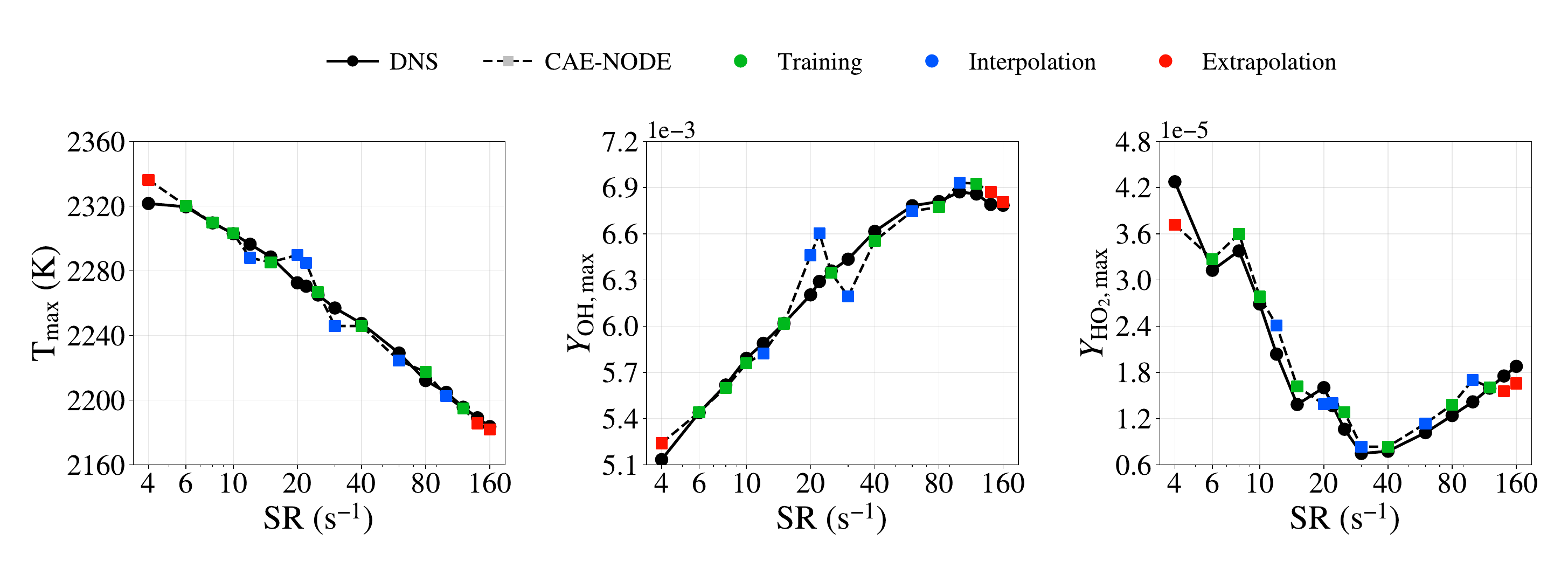}
    \caption{Maximum value of selected variables along the centerline at the last time step (10 ms), DNS vs CAE-NODE. X-axis is plotted on a log-scale. Red, green, and blue markers refer to extrapolation, training, and interpolation strain rates, respectively.}
    \label{fig:max_centerline_10ms}
\end{figure*}

Figure~\ref{fig:max_centerline_10ms} compares the maximum temperature and the maximum mass fractions of \ce{HO2} and \ce{OH} along the centerline at 10\,ms, representing
the established 2D counterflow diffusion flame, across the full range of strain rates. The model accurately tracks all three trends, and confirms its ability to capture the relationship between the flame dynamics and the governing strain rate. 

\begin{table}[htpb!]
\centering
\caption{RMSE and RFE of CAE-NODE predictions for the validation cases, SR = 4, 60 and 160 s$^{-1}$. Chemical species name represents mass fraction for that species.}
\setlength{\tabcolsep}{4pt}
\footnotesize
\begin{tabular}{lcccc}
\toprule
 & \multicolumn{2}{c}{RMSE} & \multicolumn{2}{c}{RFE [\%]} \\
\cmidrule(lr){2-3} \cmidrule(lr){4-5}
T/Y$_k$ & space & manifold & space & manifold \\
\midrule
\multicolumn{5}{c}{SR = 4 s$^{-1}$} \\
\midrule
T [K]   & $1.16\times10^{2}$  & 24.4                & 9.97   & 2.13  \\
CH$_4$  & $3.40\times10^{-2}$ & $1.74\times10^{-3}$ & 7.92   & 0.41  \\
O$_2$   & $1.76\times10^{-2}$ & $4.76\times10^{-3}$ & 11.94  & 3.27  \\
H$_2$O  & $8.74\times10^{-3}$ & $1.93\times10^{-3}$ & 17.22  & 3.94  \\
OH      & $5.39\times10^{-4}$ & $5.66\times10^{-5}$ & 39.47  & 4.47  \\
HO$_2$  & $1.30\times10^{-5}$ & $2.20\times10^{-6}$ & 103.16 & 25.63 \\
\midrule
\multicolumn{5}{c}{SR = 60 s$^{-1}$} \\
\midrule
T [K]   & 15.1                & 14.8                & 1.71  & 1.70  \\
CH$_4$  & $5.68\times10^{-3}$ & $1.28\times10^{-3}$ & 0.94  & 0.21  \\
O$_2$   & $2.62\times10^{-3}$ & $2.49\times10^{-3}$ & 1.64  & 1.58  \\
H$_2$O  & $1.27\times10^{-3}$ & $1.52\times10^{-3}$ & 3.77  & 4.67  \\
OH      & $6.38\times10^{-5}$ & $2.88\times10^{-5}$ & 6.55  & 3.14  \\
HO$_2$  & $2.27\times10^{-6}$ & $4.15\times10^{-7}$ & 38.62 & 12.55 \\
\midrule
\multicolumn{5}{c}{SR = 160 s$^{-1}$} \\
\midrule
T [K]   & 22.4                & 18.9                & 2.74  & 2.33  \\
CH$_4$  & $1.44\times10^{-2}$ & $1.72\times10^{-3}$ & 2.03  & 0.24  \\
O$_2$   & $3.28\times10^{-3}$ & $2.66\times10^{-3}$ & 2.25  & 1.85  \\
H$_2$O  & $2.24\times10^{-3}$ & $2.22\times10^{-3}$ & 7.75  & 7.84  \\
OH      & $1.38\times10^{-4}$ & $4.79\times10^{-5}$ & 17.04 & 6.21  \\
HO$_2$  & $2.19\times10^{-6}$ & $6.05\times10^{-7}$ & 48.81 & 21.17 \\
\bottomrule
\end{tabular}
\label{tab:relative_errors_unseen}
\end{table}

\subsection{Latent Space} \label{sec:latent}
\begin{figure}
    \centering
     \includegraphics[width=0.9\linewidth]{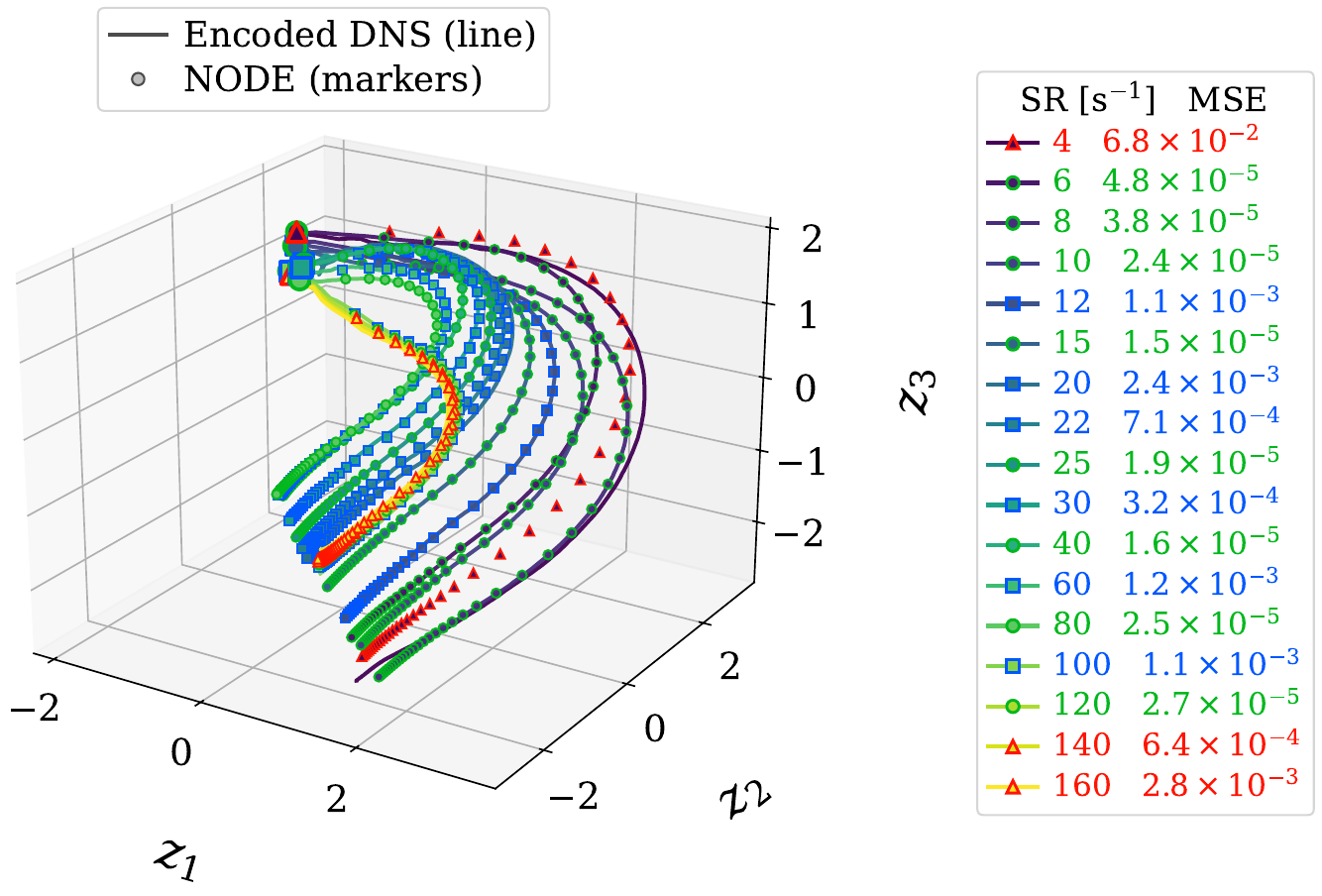}
    \caption{Latent space phase portrait. Red, green, and blue markers refer to extrapolation, training, and interpolation strain rates, respectively.}
    \label{fig:latent_all}
\end{figure}

\begin{figure*}[ht!]
    \begin{subfigure}{1\textwidth}
        \centering
        \includegraphics[width=0.7\linewidth]{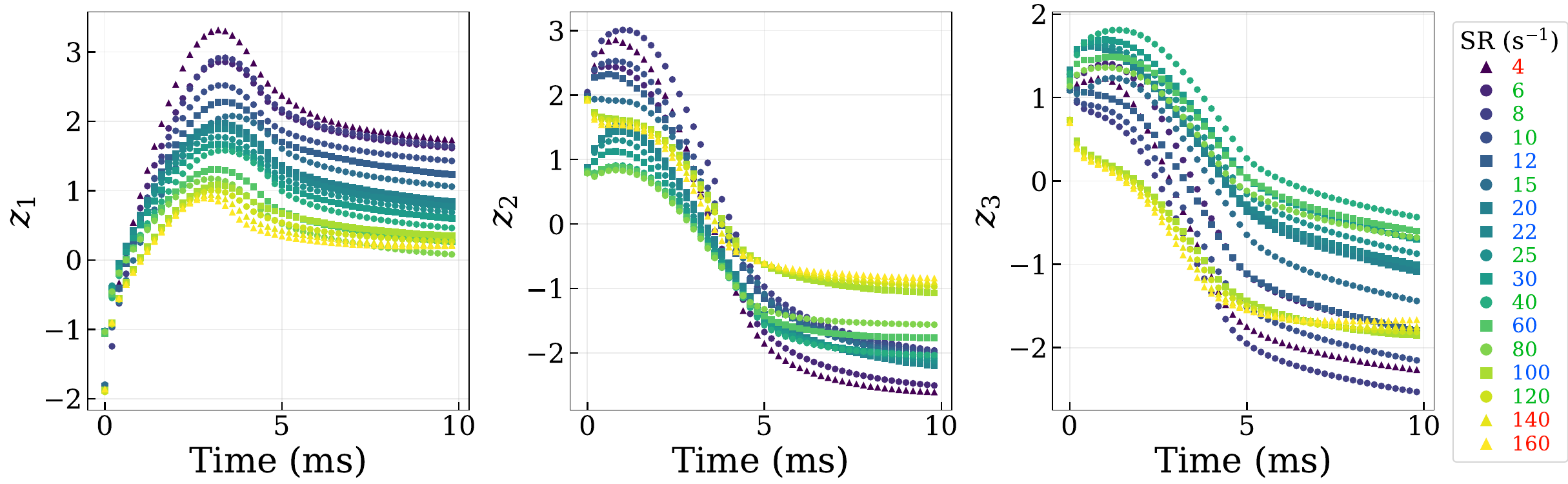}
        \caption{Latent Components over time.}
        \label{fig:latent_components}
    \end{subfigure}\vfill
    \begin{subfigure}{1\textwidth}
        \centering
        \includegraphics[width=0.7\linewidth]{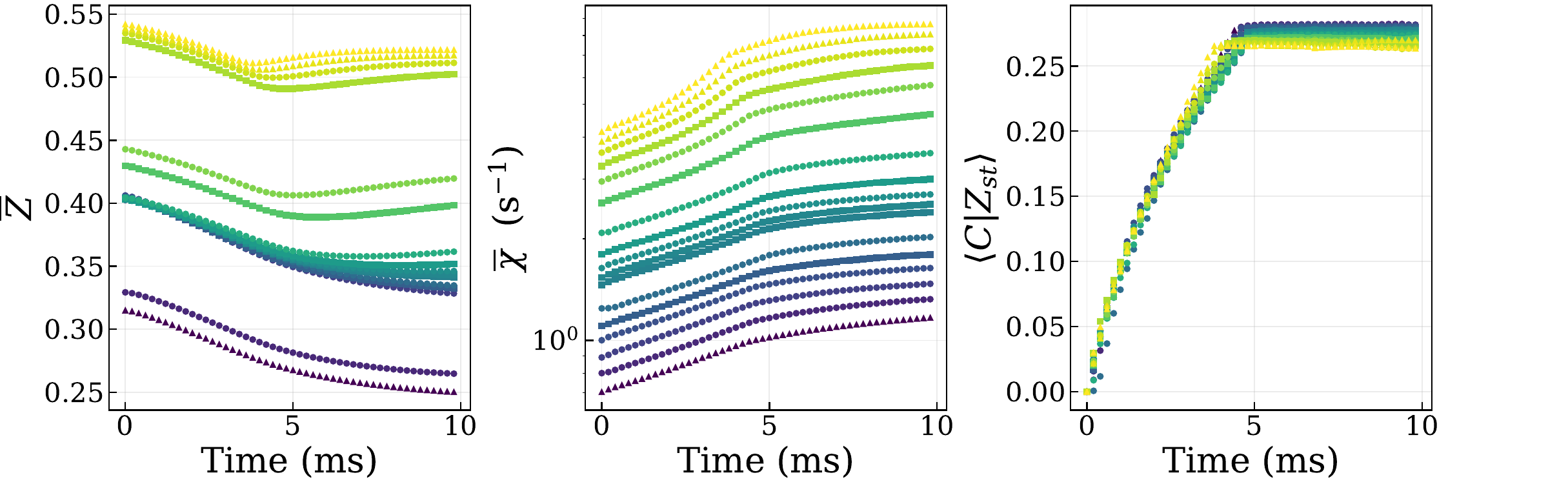}
        \caption{Global flame-state descriptors ($\overline{Z}$, $\overline{\chi}$, $\langle C | Z_{st} \rangle$) over time.}
        \label{fig:state_vars_vs_time}
    \end{subfigure}\vfill
    \vspace{5pt}
    \begin{subfigure}{1\textwidth}
        \centering
        \includegraphics[width=0.7\linewidth]{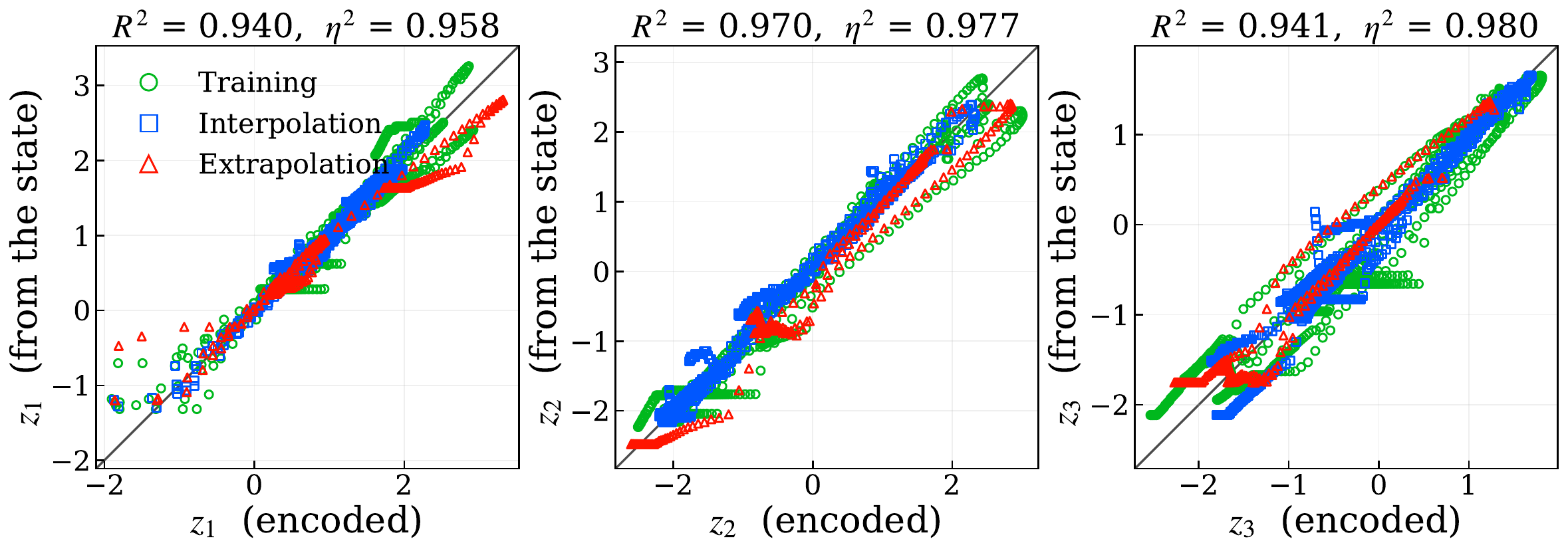}
        \caption{Correlation between latent variables predicted from the global flame-state descriptors ($\overline{Z}$, $\overline{\chi}$, $\langle C | Z_{st} \rangle$) vs the encoded latent variables.}
        \label{fig:parity_state_vs_latent}
    \end{subfigure}\vfill
    \caption{Latent variables and the global flame-state descriptors  .}
    \label{fig:flame-state_and_latent}
\end{figure*}

Figures~\ref{fig:latent_all} and~\ref{fig:latent_components} show the learned latent representation. In the phase portrait, the trajectories originate from a compact but distinct set of initial states, reflecting the different mixing conditions established at each strain rate before the hot spot is introduced, and fan out smoothly and monotonically with the strain rate without folding or intersecting one another. This arrangement is not imposed by the architecture but emerges from the joint training. The projection loss (Eq.~\eqref{eq:L3}) requires the encoded snapshots to lie along a trajectory that a single smooth vector field $h_\theta$, conditioned on $\log(\mathrm{SR})$, can reproduce by integration. Since the encoder is updated together with the NODE, a representation that is not consistent with such a field cannot be compensated by the NODE alone, and the encoder is instead driven towards one in which the temporal evolution is smooth and varies systematically with the strain rate. The reconstruction losses (Eqs.~\eqref{eq:L1} and~\eqref{eq:G1}) prevent this from degrading into a trivially integrable representation, since the latent state must still decode to the full-resolution thermochemical fields. The resulting structure, in which the strain rate acts as a continuous deformation of a single family of trajectories, leads to the accurate interpolation between the trained strain rates as well as the accurate extrapolation beyond them. In the individual component panels, the curves do cross one another. These crossings are a consequence of viewing one-dimensional projections of a three-dimensional trajectory; since the latent dynamics are conditioned on $\log(\mathrm{SR})$, the relevant state is the augmented $(z, \log\mathrm{SR})$ rather than any individual latent coordinate. 

For the trained strain rates, the NODE prediction is indistinguishable from the encoded truth, with latent MSE values between $1.5\times10^{-5}$ and $4.8\times10^{-5}$. The interpolated cases behave in the same manner: the predicted markers follow the encoded trajectories over the full $10$~ms window, at MSE values of $3.2\times10^{-4}$ to $2.4\times10^{-3}$. Although this is one to two orders of magnitude above the trained conditions, the trajectories remain qualitatively identical to the encoded DNS trajectories, and no interpolated case departs from the manifold. The same holds for extrapolation for higher strain rates, where $\mathrm{SR}=140$ and $160$ reach $6.4\times10^{-4}$ and $2.8\times10^{-3}$, respectively; the trajectories are closer together at the higher strain rates, so stepping beyond the largest trained condition requires only a small displacement in the latent space. The single exception is $\mathrm{SR}=4$ s$^{-1}$, where the predicted markers visibly detach from the encoded trajectory and the MSE rises to $6.8\times10^{-2}$, roughly three orders of magnitude above the trained cases. At the low-strain-rate end, the trajectories separate rapidly compared to higher-strain-rate cases, so extrapolating below the smallest trained condition corresponds to a much larger displacement in the latent space. However, in the thermochemical state space the predicted latent trajectory is still consistent and accurate, and the displacement in the latent space manifests as a displacement in the physical space compared to the DNS. This is consistent with the elevated RMSE${_\text{space}}$, while the RFE${_\text{manifold}}$ error remains good for $\mathrm{SR}=4$ s$^{-1}$ in Table~\ref{tab:relative_errors_unseen}.

Figure~\ref{fig:parity_state_vs_latent} evaluates the physical interpretability of the unsupervised latent space by assessing whether the latent variables can be
predicted from a reduced set of global flame-state descriptors. This relationship is quantified using $K$-nearest-neighbours (KNN) regression in a
three-dimensional feature space consisting of the domain-averaged mixture fraction, $\overline{Z}$; the domain-averaged scalar
dissipation rate, $\overline{\chi}$; and the conditional average of the progress variable at the stoichiometric surface, $\langle C | Z_{st} \rangle$. KNN regression is used because it assumes no functional form for the mapping. Each latent state is estimated as an unweighted average of the latent states of the $K$ nearest snapshots in the flame-state descriptor space, analogous to a tabulated look-up that interpolates between stored entries. A high prediction accuracy therefore indicates that snapshots which are close in the flame-state descriptor space are also close in the latent space, that is, that the descriptors determine the latent state, rather than that a particular regression model happens to fit the data.

These three quantities are selected because they characterize complementary aspects of the flame state. The domain-averaged mixture fraction
$\overline{Z}$ varies inversely with the thickness of the mixing layer established by the imposed strain rate. After ignition, as the flame kernel
expands, $\overline{Z}$ decreases and reaches a minimum, after which it recovers for the higher strain rates as the mixing layer flattens into the
steady-state diffusion flame, while it continues to decrease at the lower strain rates where no such flattening occurs. The domain-averaged scalar
dissipation rate $\overline{\chi}$ increases with strain rate and, unlike $\overline{Z}$, grows monotonically as the reaction progresses. The
conditional progress variable at stoichiometry $\langle C|Z_{st}\rangle$ measures how far reaction has advanced on the flame surface and therefore
tracks the transient from ignition to the established diffusion flame. Figure~\ref{fig:state_vars_vs_time} shows how these variables differ from each other.
$\overline{Z}$ and $\overline{\chi}$ separate into well-ordered bands with strain rate while varying comparatively little in time, whereas
$\langle C|Z_{st}\rangle$ collapses onto a single curve across all strain rates while monotonically increasing with time. The three descriptors
together therefore locate a snapshot in both the operating condition (strain rate) and time. To identify these set of variables all two- and three-variable combinations of six candidate quantities, $\overline{Z}$, $\overline{C}$, $\overline{\chi}$, $\overline{T}$, $\langle C|Z_{st}\rangle$ and $\langle \chi|Z_{st}\rangle$, were evaluated with the same KNN procedure and ranked by the mean $\eta^2$ across the three latent dimensions. The combination used ranked the highest.

Adjacent snapshots along a given trajectory are strongly correlated, so snapshots from the same strain rate are excluded from the KNN regression and
the neighbors are drawn exclusively from different strain-rate trajectories. Each latent state is thus reconstructed from similar physical
states realized at other strain rates rather than from its own highly correlated neighbors. The regression then answers the practical question of how
well the latent variables of an unseen strain rate could be predicted from the three flame-state descriptors and the latent variables of a known set of simulations.

The predictions (Fig.~\ref{fig:parity_state_vs_latent}) closely follow the 1:1 line across the training, interpolation, and extrapolation conditions. This agreement is quantified using the coefficient of
determination, $R^2$, which evaluates the accuracy of the KNN prediction, and the correlation ratio, $\eta^2$, which measures the residual spread of the
latent variable within each neighbourhood and therefore the fraction of its variance explained by the selected descriptors. The reconstruction uses
$k = 12$ neighbours, and both measures were found to be insensitive to this choice over the range $k = 3$ to $40$.

The three latent dimensions achieve $R^2$ values of $0.940$, $0.970$ and $0.941$, and $\eta^2$ values of $0.958$, $0.977$ and $0.980$ for $z_1$, $z_2$
and $z_3$, respectively, with $\eta^2 = 0.971$ for the combined latent vector. The three global flame-state descriptors ($\overline{Z}$, $\overline{\chi}$,
$\langle C|Z_{st}\rangle$) therefore account for roughly 97\% of the variance of the latent space. $R^2$ being slightly below $\eta^2$ follows from, for a given snapshot, excluding the states from their own trajectory. The KNN utilizes neighbors from adjacent trajectories that sit at a marginally displaced position on the latent manifold. This displacement causes a slight error that reflects in $R^2$, but does not affect $\eta^2$. The small gap between the two thus indicates the magnitude of the strain-rate dependence not captured by the descriptors, which is just a few percent of the latent variance.

Together, these results indicate that, despite the absence of any supervision, the CAE-NODE framework has learned a latent manifold informed by the physics of
the transient 2D counterflow flame, dependent on the strain rate. The manifold is recovered to within a few percent of its variance by three
quantities based on $C$ and $Z$, the same variables used for reduced-order models in the flamelet-generated manifold framework. The network is therefore
not merely fitting the temporal evolution of the training simulations, but organizing its internal representation along coordinates that carry physical
meaning, and it does so without any explicit "physics-informed" loss function or knowledge of these descriptors during the training process. 

This also accounts for the accuracy of the model outside the trained range. The descriptors vary smoothly and monotonically with the strain rate, and the latent space captures this variation to a high accuracy.

\subsection{Stiffness Reduction \& Computational Speedup}

\begin{figure}[!h]
\centering
\includegraphics[width=0.8\linewidth]{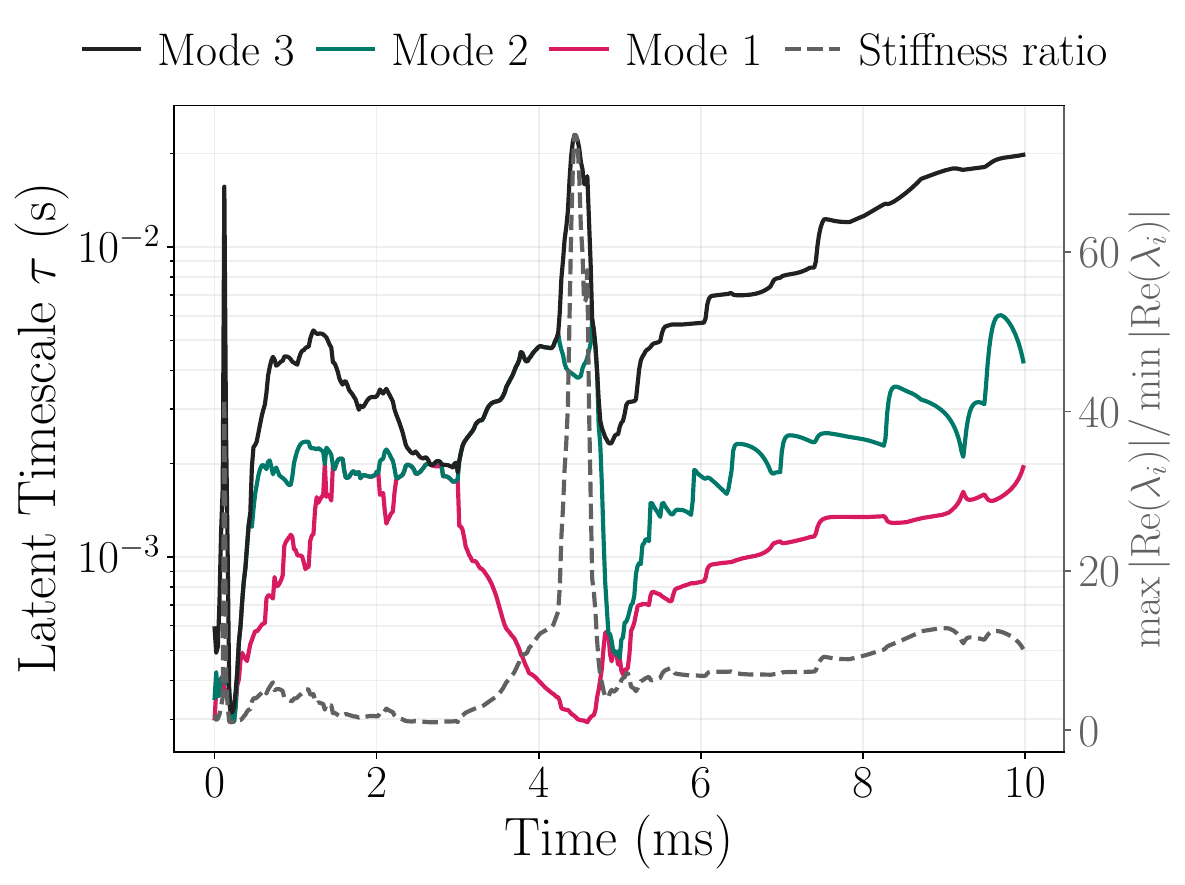}
\caption{\footnotesize Latent timescales over time for SR = 40 s$^{-1}$.}
\label{fig:latent_timescales_adaptive}
\end{figure}

The transient flame considered here spans both the ignition and propagation
phases, and therefore covers multiple combustion regimes. Projecting from the
high-dimensional physical space ($256\times256\times21$) onto a
three-dimensional latent space substantially reduces the temporal stiffness of
the resulting dynamical system. A timescale analysis of the latent vector field
(Figure~\ref{fig:latent_timescales_adaptive}) gives a fastest latent timescale
of 3$\times10^{-4}$~s, several orders of magnitude larger than the
characteristic chemical timescales that limit time integration in reactive CFD,
which range from $\mathcal{O}(10^{-8})$ to $\mathcal{O}(10^{-6})$~s. Defining
the stiffness ratio as
\begin{equation}
    S(t) = \frac{\max_i \left| \mathrm{Re}\big(\lambda_i(t)\big) \right|}
                {\min_i \left| \mathrm{Re}\big(\lambda_i(t)\big) \right|},
    \label{eq:stiffness_ratio}
\end{equation}
where the $\lambda_i$ are the eigenvalues of the Jacobian $\partial h(z) / \partial z$ of the latent vector field evaluated along the trajectory at fixed strain
rate, $S$ peaks at approximately $70$ and remains below $10$ over most of the integration. For comparison, the chemical mechanism in physical space exhibits
$S = \mathcal{O}(10^{8})$ or larger stiffness ratios, due to both fast and slow time scales existing together. The manifold learned by CAE-NODE is therefore non-stiff in
practice and can be advanced with an explicit integrator at large time steps.

\begin{figure}[!h]
\centering
\includegraphics[width=0.83\linewidth]{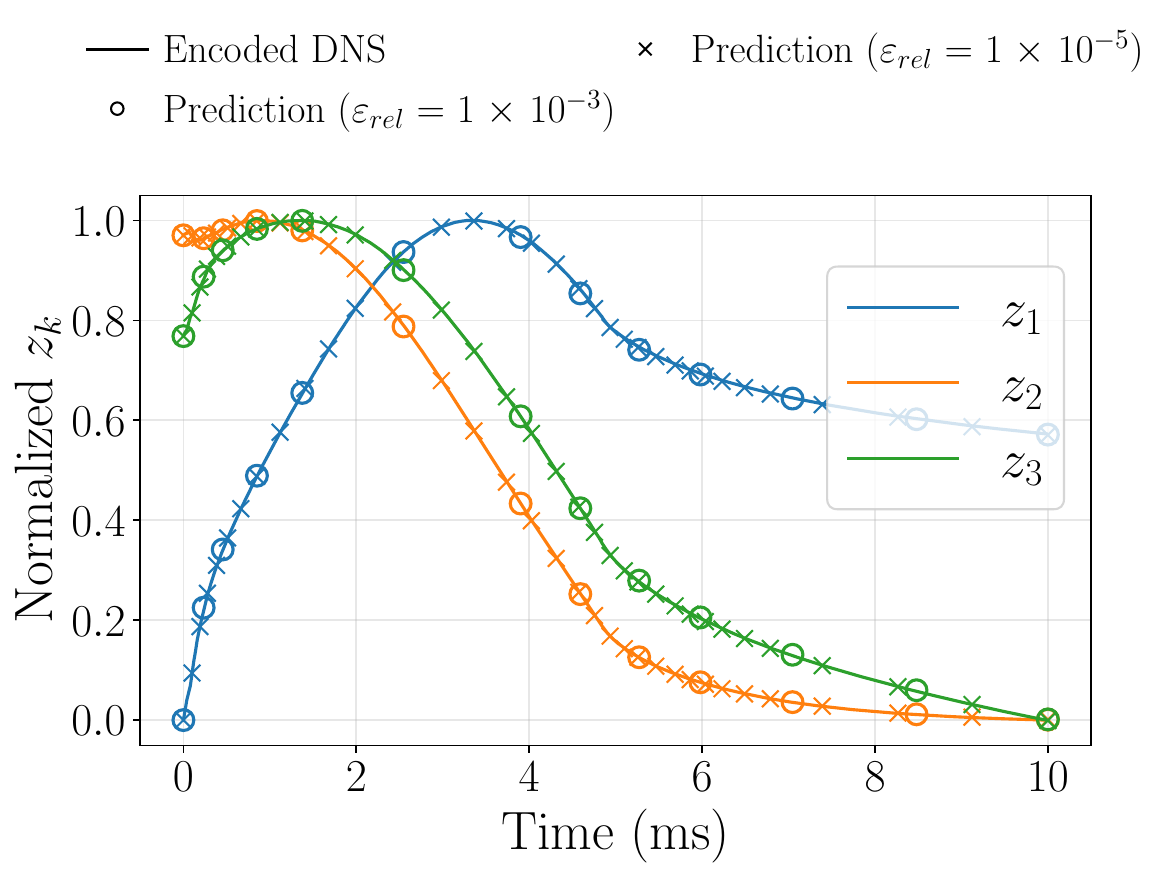}
\caption{\footnotesize Normalized latent variables over time for the SR = 40 s$^{-1}$, with adaptive time stepping and $\varepsilon_{rel} = [1\times10^{-3}, 1\times10^{-5}]$.}
\label{fig:latent_variables_adaptive}
\end{figure}

Figure~\ref{fig:latent_variables_adaptive} compares the evolution of the three latent variables for two CAE-NODE time integrations using the same network and absolute
tolerance ($1\times10^{-16}$) but relative tolerances of $1\times10^{-3}$ and $1\times10^{-5}$, integrated with an explicit adaptive Runge-Kutta (RK45)
scheme. The trajectories are nearly identical. For these tolerances, only 13 and 34 adaptive steps, respectively, are required to advance the flame from 0 to 10~ms. Relative to the 10{,}000 uniform steps used in the CFD simulation, this is a reduction of approximately $770\times$ and $290\times$. 

The corresponding CFD simulation requires approximately $8.3\times10^{4}$~CPU core-seconds on an AMD EPYC 9354. The CAE-NODE prediction over the same
interval, including decoding to the full $256\times256\times21$ field at 500 uniform output times, requires about 31~CPU core-seconds on an Intel
i5-13600KF and approximately 1~s of wall-clock time on a single NVIDIA RTX~4090 GPU, an overall reduction of roughly three orders of magnitude. The hardware differs
between the two measurements, so these figures should be read as indicative rather than as a controlled benchmark.

\subsection{Comparison with PC-transport}

In this section, a comparison is shown between the CAE-NODE approach and the multi-linear data-based PC-transport approach \cite{sutherland2009combustion} based on Principal Component Analysis \cite{PCA_Jolliffe}. In this model, transport equations are solved for a reduced number of optimal variables containing most of the variance in the training data set. A transport equation for those new variables, called PC scores, is obtained by projecting the species transport onto the PCA basis matrix $\mathbf{A}_{\mathrm{q}}$ \cite{sutherland2009combustion}:
\begin{equation}
\frac{\partial}{\partial t}\left(\rho\mathbf{z}\right)+\nabla\cdot\mathbf{C_{z}}=\nabla\cdot\mathbf{J_{z}}+\mathbf{S_{z}}\label{eq:PC transport}
\end{equation}
where  $\rho$ is the density, $\mathbf{C_{z}}$, $\mathbf{J_{z}}$ and $\mathbf{S_{z}}$ are the convective flux, diffusive flux and chemical source terms for the scores, respectively. $\mathbf{C_{z}}$ is defined as $\mathbf{C_{z}}=\rho\mathbf{u}\mathbf{z}$ where $\mathbf{u}$ is the velocity vector, 
$\mathbf{S_{z}}$ is defined as $\mathbf{S_{z}}=\mathbf{S_{y}\,A_{\mathrm{q}}}$
with $\mathbf{S_{y}}$ being the species chemical source terms, and $\mathbf{J_{z}}=\rho\mathbf{D_{z}}\nabla\mathbf{z}$
where the score diffusion matrix $\mathbf{D_{z}}=\mathbf{A}_{\mathrm{q}}^{\mathrm{T}}\mathbf{D_{y}}\mathbf{A}_{\mathrm{q}}$
and $\mathbf{D_{y}}$ is the diagonal matrix of the species diffusion coefficients.

A PCA was performed using the DNS data set, combining all strain rate cases and all temporal snapshots. The analysis was done using all the $20$ chemical species in the mechanism, which then resulted in $20$ components (PCs) for the full PCA basis. The PC-transport simulations were carried out using a custom implementation within OpenFOAM.

\begin{figure}[!h]
\centering
\includegraphics[width=1.0\linewidth]{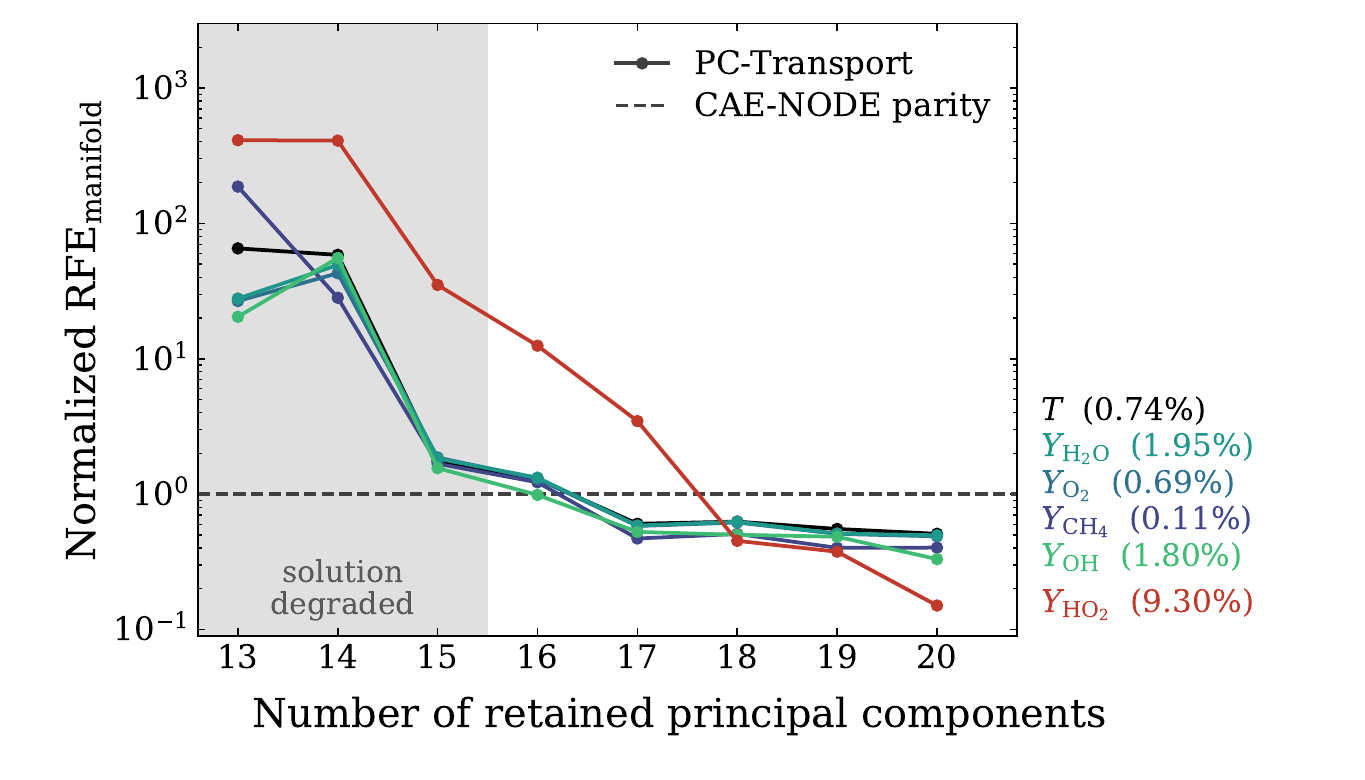}
\caption{\footnotesize A posteriori  RFE${_\text{manifold}}$ for selected thermochemical variables vs number of PCs retained for SR = 40 s$^{-1}$ normalized by  RFE${_\text{manifold}}$ of CAE-NODE.}
\label{fig:pc_vs_caenode}
\end{figure}

Figure \ref{fig:pc_vs_caenode} shows the normalized RFE${_\text{manifold}}$ error for the SR$=40~s^{-1}$ case, comparing the CAE-NODE model (horizontal dashed line) with the PC-transport simulations (solid lines) as a function of the number of retained PCs for different thermochemical variables. It can be observed that for the PC-transport case, the error increases as the number of transported variables is reduced. Using 17 PCs provided an error comparable to the CAE-NODE. Below 16 PCs, the solution becomes unphysical. The 17 PCs case was therefore retained for the subsequent analysis. 

\begin{figure*}[ht!]
\centering
\begin{subfigure}[h]{1\textwidth}
    \centering
    \includegraphics[width=\textwidth]{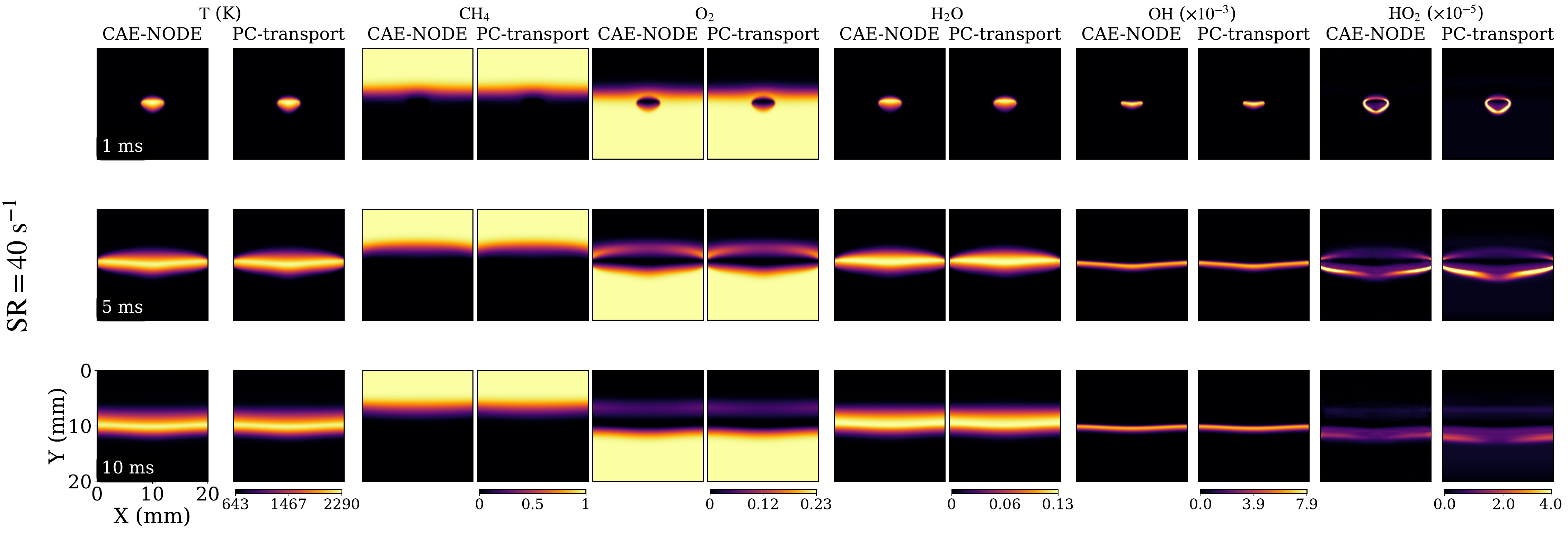}
    \caption{\footnotesize Temperature, and Y$_k$ snapshots.}
    \label{pca17_snapshots}
    
\end{subfigure}
\begin{subfigure}[h]{0.9\textwidth}
    \centering
    \includegraphics[width=\textwidth]{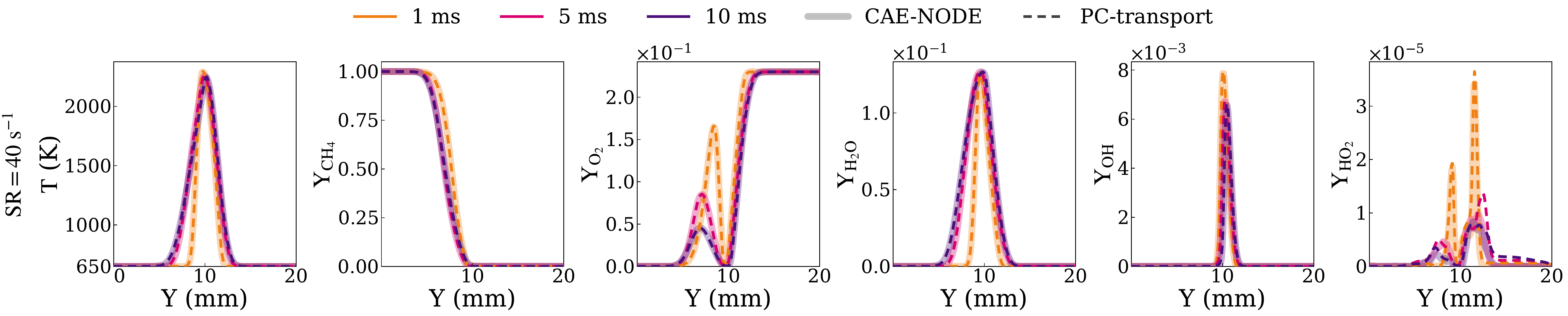}
    \caption{\footnotesize Temperature and Y$_k$ along the centerline.}
    \label{pca17_centerline}
    
\end{subfigure}
\begin{subfigure}[h]{0.9\textwidth}
    \centering
    \includegraphics[width=\textwidth]{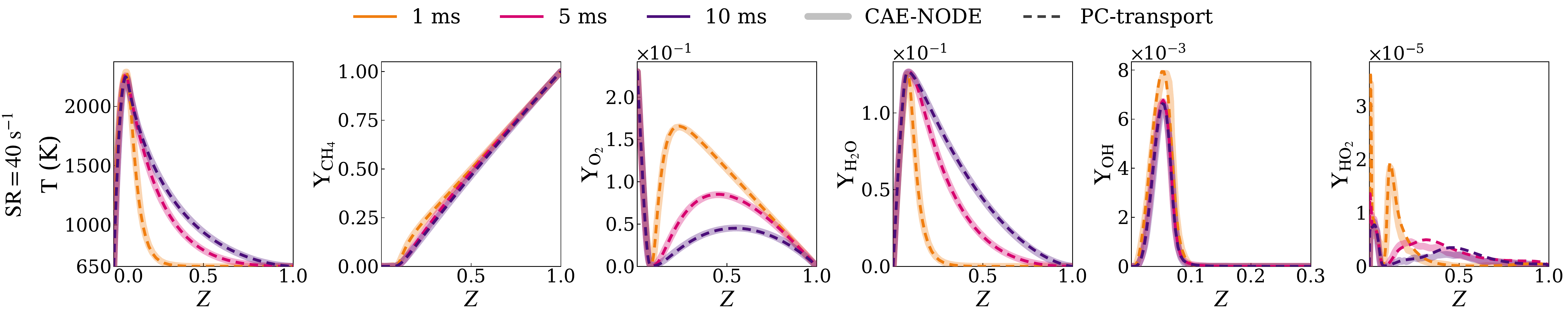}
    \caption{\footnotesize Temperature and Y$_k$  vs mixture fraction ($Z$) along the centerline.}
    \label{pca17_centerline_Zspace}
\end{subfigure}

\caption{\footnotesize Comparison of CAE-NODE vs PC-transport with 17 PCs for SR = 40 s$^{-1}$.}
\label{fig:center_combined_PCA17}
\end{figure*}

Figure \ref{pca17_snapshots} compares the CAE-NODE and PC-transport (using 17 PCs) snapshots of temperature and selected species mass fraction at 1, 5, and 10\,ms for strain rate SR~$=40\,\mathrm{s}^{-1}$ (similar to Fig.~\ref{snapshots_SR40}). The PC-transport and CAE-NODE results are very similar for the flame at all three stages, including for the curvature of the flame front at 5 ms. Minor species are also similarly represented; however, a slight discrepancy in the \ce{HO2} snapshot can be observed at 5 and 10 ms for the PC-transport case, which correlates to the \ce{HO2} error observed in Fig. \ref{fig:pc_vs_caenode} for 17 PCs.

Figures \ref{pca17_centerline} and \ref{pca17_centerline_Zspace} show a quantitative comparison of the PC-transport ROM with CAE-NODE using centerline profiles in the physical space (Fig. \ref{pca17_centerline}) and in mixture fraction space (Fig. \ref{pca17_centerline_Zspace}). Using 17 PCs, the PCA-based ROM reproduces the CAE-NODE results for all the thermochemical variables at all three instants, making both ROM results indistinguishable from each other and preserving the state-space structure of the flame.

In this study, the global version of the PC-transport model was used, where the PCA basis matrix $\mathbf{A}_{\mathrm{q}}$ is constant over the entire manifold. This explains the modest reduction obtained compared to the CAE-NODE model. Recently, Malik et al. \cite{Malik_PROCI_2025,Malik_PROCI_2026} introduced a local version of PC-transport, where the state-space manifold is first divided into separate clusters, and subsequently a local PCA basis is calculated into each one of them. Those local PC scores are then transported, which considerably improves the achievable size reduction and acceleration of the simulation. This approach could be pursued in a future work on CAE-NODE.

\section{Conclusion}
This study successfully demonstrated a novel Convolutional Autoencoder and Neural ODE (CAE-NODE) framework as a robust, data-driven reduced-order model for multi-dimensional, transient reacting flows. By applying the model to two-dimensional counterflow methane-air flames, the framework achieved a compression ratio of over 400,000, mapping high-dimensional thermochemical fields to a highly compact three-dimensional latent space. 

The results establish that the CAE-NODE architecture accurately captures the full transient evolution of the flame - from ignition through propagation to the established diffusion flame - with excellent agreement with Direct Numerical Simulation (DNS) data. Crucially, the decoder naturally preserves mass and elemental conservation principles across the mixture fraction space without requiring explicit physics-informed constraints during training. Furthermore, the model exhibits robust generalization capabilities, accurately predicting flame structures and dynamics for unseen strain rates in both interpolative and extrapolative regimes. 

Despite the unsupervised formulation of the latent manifold, the learned latent variables proved to be highly interpretable and physically meaningful. A reduced feature space consisting of the domain-averaged mixture fraction, scalar dissipation rate, and conditional progress variable was shown to account for approximately 97\% of the variance within the latent space. Finally, projecting the system into this low-dimensional manifold significantly reduced the temporal stiffness of the reactive system, enabling the use of large explicit time steps. This stiffness reduction ultimately yielded a computational speedup of approximately three orders of magnitude compared to the reference DNS, demonstrating the CAE-NODE framework's exceptional potential as a fast, accurate, and physically consistent surrogate model for complex combustion phenomena.

The CAE-NODE was also successfully validated against a well-established unsupervised ROM (PC-transport), showing its promise in the field of reacting flow simulations.

\section*{CRediT authorship contribution statement}

\textbf{Mert Yakup Baykan}: Conceptualization, Methodology (CAE-NODE), Investigation, Writing - Original Draft. 
\textbf{Weitao Liu}: Conceptualization, Methodology (DNS), Investigation, Writing - Original Draft. 
\textbf{Mohammad Rafi Malik}: Methodology (PC-transport), Writing - Original Draft. 
\textbf{Thorsten Zirwes}: Writing - Review \& Editing, Supervision. 
\textbf{Andreas Kronenburg}: Writing - Review \& Editing, Supervision. 
\textbf{Hong G. Im}: Writing - Review \& Editing, Supervision. 
\textbf{Dong-hyuk Shin}: Writing - Review \& Editing, Supervision.

\section*{Declaration of competing interest}

The authors declare that they have no known competing financial
interests or personal relationships that could have appeared to
influence the work reported in this paper.

\section*{Acknowledgments}

This work was supported by “Human Resources Program in Energy Technology” of the Korea Institute of Energy Technology Evaluation and Planning (KETEP), granted financial resource from the Ministry of Trade, Industry \& Energy, Republic of Korea (Grant: RS-2021-KP002521), and was supported by the InnoCORE program of the Ministry of Science and ICT (N10250155). Weitao Liu acknowledges financial support from the China Scholarship Council (CSC, No. 202108340023), and Prof. Andreas Kronenburg acknowledges financial support from the Deutsche Forschungsgemeinschaft (DFG, No. 462463789).


\FloatBarrier

\bibliographystyle{cnf-num}
\bibliography{cnf-refs}

\end{document}